\pdfoutput=1
\documentclass[sigconf,screen]{acmart}
\AtBeginDocument{%
  \providecommand\BibTeX{{%
    \normalfont B\kern-0.5em{\scshape i\kern-0.25em b}\kern-0.8em\TeX}}}

\usepackage[T1]{fontenc}
\usepackage[ansinew]{inputenc}
   
\usepackage{booktabs}   
\usepackage{subcaption} 
\usepackage{paralist}
\usepackage{enumitem}

\usepackage{xstring}

\usepackage[capitalise]{cleveref}
\usepackage{relsize}
\usepackage{siunitx}
\usepackage{tikz}
\usepackage{xcolor}
\usepackage{multirow}
\usepackage{pgfplots}
\usepackage{fontawesome5}
\usepackage{lipsum}
\usepackage{listings}
\usepackage{microtype}
\usepackage{pifont}
\usepackage[all]{nowidow}

\setlipsum{%
  par-before = \begingroup\color{red},
  par-after = \endgroup
}

\usetikzlibrary{arrows}
\usetikzlibrary{chains}
\usetikzlibrary{calc}
\usetikzlibrary{trees}
\usetikzlibrary{fit}
\usetikzlibrary{matrix}
\usetikzlibrary{positioning}
\usetikzlibrary{quotes}
\usetikzlibrary{shapes}
\usetikzlibrary{patterns, patterns.meta}
\usetikzlibrary{decorations, decorations.pathmorphing, decorations.pathreplacing}
\usetikzlibrary{pgfplots.groupplots}
\usetikzlibrary{shapes.geometric, shapes.symbols}

\pgfplotsset{compat=newest}%

\definecolor{col0o}{RGB}{0,146,146}
\colorlet{col0}{col0o!80}
\definecolor{col1o}{RGB}{182,109,255}
\colorlet{col1}{col1o!70}
\definecolor{col2}{RGB}{219,209,0}
\definecolor{col3}{RGB}{255,182,119}
\definecolor{col8}{RGB}{182,255,219}
\definecolor{col5}{RGB}{36,255,36}
\definecolor{col6}{RGB}{182,219,255}
\definecolor{col7}{RGB}{255,109,182}
\definecolor{col18}{RGB}{255,255,109}
\definecolor{col9}{RGB}{73,0,146}
\definecolor{col10}{RGB}{146,73,0}
\definecolor{col11}{RGB}{146,0,0}
\definecolor{col12}{RGB}{0,109,219}
\definecolor{col13}{RGB}{0,73,73}
\definecolor{col14}{RGB}{73,0,73}
\definecolor{col15}{RGB}{73,73,0}
\definecolor{col16o}{RGB}{255,36,36}
\colorlet{col16}{col16o!80}
\definecolor{col17o}{RGB}{36,36,255}
\colorlet{col17}{col17o!50}
\definecolor{col4}{RGB}{255,182,219}
\definecolor{col19}{RGB}{109,182,255}
\colorlet{commentcolor}{black!80}

\pgfplotscreateplotcyclelist{mycolorlist}{%
fill=col0\\%
fill=col1\\%
fill=col2\\%
fill=col11\\%
fill=col4\\%
fill=col13\\%
fill=col7\\%
fill=col12\\%
fill=col9\\%
fill=col10\\%
fill=col3\\%
fill=col5\\%
fill=col14\\%
fill=col15\\%
fill=col16\\%
fill=col17\\%
}
\colorlet{commentcolor}{black!80}
\colorlet{pragmacolor}{col10}

\tikzset{
  >=stealth,
  thread/.style={->,decorate,
    decoration={snake,amplitude=.4mm, segment length=2mm, post length=1mm},
    thick},
  workerthread/.style={
    black!50,
    dash pattern=on 3pt off 1.0pt,
    thread,
  },
  sm/.style={
    rectangle,
    rounded corners,
    fill=white,
    draw=black, very thick,
    inner sep = 2mm,
    minimum width=3em,
    minimum height=3em,
    text centered},
  mynode/.style={
    rectangle,
    rounded corners,
    draw=black, very thick,
    inner sep = 2mm,
    minimum height=3em,
    text centered},
  line/.style={draw, thick, <-},
  element/.style={
    tape,
    top color=white,
    bottom color=blue!50!black!60!,
    minimum width=8em,
    draw=blue!40!black!90, very thick,
    text width=10em,
    minimum height=3.5em,
    text centered},
  every join/.style={->, ultra thick,shorten >=1pt},
  tuborg/.style={decorate, ultra thick,white,text=black},
  midnode/.style={midway, above=2pt},
  midnodebelow/.style={midway, below=2pt},
  mybrace/.style={decorate,decoration={brace,aspect=#1,amplitude=10pt}},
  device/.style={
    rectangle,
    rounded corners,
    fill=white,
    draw=black, very thick,
    inner sep = 2mm,
    minimum width=3.5em,
    minimum height=4em,
    text centered},
  host/.style={
    rectangle,
    fill=white,
    draw=black, thick,
    inner sep = 1mm,
    minimum width=1em,
    minimum height=1em,
    text centered},
}
\pgfplotsset{
    ylabel style = {align=center,at={(-0.08,0.5)}},
    ymajorgrids=true,
    grid style=dashed,
}
\newdimen\XCoordA
\newdimen\YCoordA
\newcommand*{\ExtractCoordinateA}[1]{\path (#1); \pgfgetlastxy{\XCoordA}{\YCoordA};}%
\newdimen\XCoordB
\newdimen\YCoordB
\newcommand*{\ExtractCoordinateB}[1]{\path (#1); \pgfgetlastxy{\XCoordB}{\YCoordB};}%

\def\ContinueLineNumber{\lstset{firstnumber=last}}

\begin{document}

\title{GPU First --- Execution of Legacy CPU Codes on GPUs}
\author{Shilei Tian}
\email{shilei.tian@stonybrook.edu}
\orcid{0000-0001-6468-6839}
\affiliation{%
  \institution{Stony Brook University}
  \streetaddress{100 Nicolls Road}
  \city{Stony Brook}
  \state{NY}
  \country{USA}
  \postcode{11794}
}

\author{Tom Scogland}
\email{scogland1@llnl.gov}
\orcid{0000-0001-7234-5743}
\affiliation{%
  \institution{Lawrence Livermore National Laboratory}
  \streetaddress{7000 East Avenue}
  \city{Livermore}
  \state{CA}
  \postcode{94550}
  \country{USA}
}

\author{Barbara Chapman}
\email{barbara.chapman@stonybrook.edu}
\orcid{0000-0001-8449-8579}
\affiliation{%
  \institution{Stony Brook University}
  \streetaddress{100 Nicolls Road}
  \city{Stony Brook}
  \state{NY}
  \postcode{11794}
  \country{USA}
}

\author{Johannes Doerfert}
\email{jdoerfert@llnl.gov}
\orcid{0000-0001-7870-8963}
\affiliation{%
  \institution{Lawrence Livermore National Laboratory}
  \streetaddress{7000 East Avenue}
  \city{Livermore}
  \state{CA}
  \postcode{94550}
  \country{USA}
}

\renewcommand{\shortauthors}{S. Tian, T. Scogland, et al.}

\begin{abstract}




Utilizing GPUs is critical for high performance on heterogeneous systems.
However, leveraging the full potential of GPUs for accelerating legacy CPU applications can be a challenging task for developers.
The porting process requires identifying code regions amenable to acceleration, managing distinct memories, synchronizing host and device execution, and handling library functions that may not be directly executable on the device.
This complexity makes it challenging for non-experts to leverage GPUs effectively, or even to start offloading parts of a large legacy application.

In this paper, we propose a novel compilation scheme called ``\emph{GPU First}'' that automatically compiles legacy CPU applications directly for GPUs without any modification of the application source.
Library calls inside the application are either resolved through our partial \texttt{libc} GPU implementation or via automatically generated remote procedure calls to the host.
Our approach simplifies the task of identifying code regions amenable to acceleration and enables rapid testing of code modifications on actual GPU hardware in order to guide porting efforts.

Our evaluation on two HPC proxy applications with OpenMP CPU and GPU parallelism, four micro benchmarks with originally GPU only parallelism, as well as three benchmarks from the SPEC OMP 2012 suite featuring hand-optimized OpenMP CPU parallelism showcases the simplicity of porting host applications to the GPU.
For existing parallel loops, we often match the performance of corresponding manually offloaded kernels, with up to $14.36\times$ speedup on the GPU, validating that our \emph{GPU First} methodology can effectively guide porting efforts of large legacy applications.

\end{abstract}





\maketitle

\section{Introduction}
\label{sec:introduction}

In today's era of high-performance computing, GPUs have emerged as the most popular solution for accelerating compute-intensive workloads due to their massive parallelism and high memory bandwidth. 
However, harnessing the full potential of GPUs can be challenging, especially for legacy CPU applications that were not designed with GPU acceleration in mind.
Porting such applications to the GPU can be time-consuming, error-prone, and usually requires significant development effort.
One needs to identify code regions amenable to acceleration, manage distinct memories, synchronize host and device execution, and handle library functions that are not executable on the device.
These tasks increase the complexity of porting and may require significant re-architecting efforts, making it difficult for non-experts to leverage GPUs for performance gains or even to initiate offloading any part of a large legacy application.


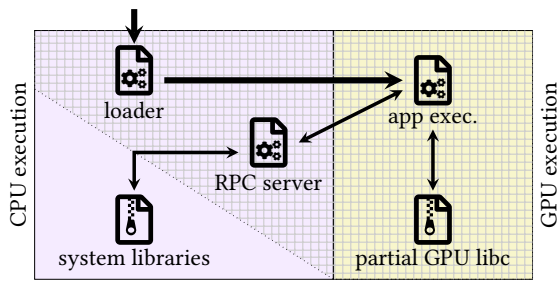
\begin{figure}[b]
    \vspace{-3mm}
    \centering
    \resizebox{0.9\linewidth}{!}{ 
      \begin{tikzpicture}[node distance=10mm]

%

\draw[fill=col1!20]
    (0.0, 0.0) rectangle (6.0, -5.0);
\draw[fill=col2!20]
    (6.0, 0.0) rectangle (10.0, -5.0);
\draw[pattern={grid},pattern color=gray!40,draw=black,dotted]
    (0.0, 0.0) -- (0.0, -1.175) -- (6.0, -5.0) -- (10, -5) -- (10, 0) -- (0,0);

\node[scale=1.5, anchor=south, rotate=90] (CPU_txt) at (0, -2.5) {CPU execution};
\node[scale=1.5, anchor=north, rotate=90] (GPU_txt) at (10, -2.5) {GPU execution};

\node[scale=2.0] (loader) at (2.0, -0.8) {\huge\faFile[regular]{}};
\node[scale=0.8,yshift=-2mm] (loader_gears) at (loader) {\huge\faCogs{}};
\node[scale=1.5,yshift=-0.5mm] (loader_txt) at (loader.south) {loader};

\node[scale=2.0] (rpc_host) at (4.7, -2.25) {\huge\faFile[regular]{}};
\node[scale=0.8,yshift=-2mm] (rpc_host_gears) at (rpc_host) {\huge\faCogs{}};
\node[scale=1.5,yshift=-0.5mm] (rpc_host_txt) at (rpc_host.south) {RPC server};

\node[scale=2.0] (library) at (2.0, -3.8) {\huge\faFileArchive[regular]{}};
\node[scale=1.5,yshift=-0.5mm] (library_txt) at (library.south) {system libraries};

\node[scale=2.0] (app) at (8.0, -1.0) {\huge\faFile[regular]{}};
\node[scale=0.8,yshift=-2mm] (device_exe_gears) at (app) {\huge\faCogs{}};
\node[scale=1.5,yshift=-0.5mm] (app_txt) at (app.south) {app exec.};

\node[scale=2.0] (gpu_library) at (8.0, -3.8) {\huge\faFileArchive[regular]{}};
\node[scale=1.5,yshift=-0.5mm] (gpu_library_txt) at (gpu_library.south) {partial GPU libc};

\ExtractCoordinateA{loader.east}
\ExtractCoordinateB{app.west}
\draw[line width=3.5pt, ->] ($(\XCoordA, \YCoordB) + (0, 0.0)$) -- ($(app.west) + (0, 0)$);
\draw[line width=3.5pt, ->] ($(loader.north) + (0, 0.5)$) -- ($(loader.north) + (0, -0.2)$);
\draw[line width=2.0pt, <->] ($(app.west) + (0, -0.2)$) -- ($(rpc_host.east) + (0, 0)$);
\draw[line width=2.0pt, <->] ($(app_txt.south) + (0, 0.15)$) -- ($(gpu_library.north) + (0, -0.15)$);

\ExtractCoordinateA{rpc_host.west}
\ExtractCoordinateB{library.north}
\draw[line width=2.0pt, <->] ($(rpc_host.west) + (0, -0.2)$) -- ($(\XCoordB, \YCoordA) + (0, -0.2)$) -- ($(library.north) + (0, -0.15)$);

\end{tikzpicture}
    }
     \vspace{-3mm}
    \caption{
    Bird's-eye view of the \emph{GPU First} methodology.
    All components on a grid background are provided or generated by the approach. 
    The loader is the entry point for the operating system and responsible to setup the environment on the device.
    The application executable (top right) is produced from the unmodified legacy source code but runs on the GPU.
    A partial \texttt{libc} GPU implementation provides relatively fast device side runtime calls while other library calls are translated into remote procedure calls (RPCs).
    Our RPC scheme will also orchestrate memory movement for arguments and underlying objects, forward the calls to existing system libraries, and return the result to the application thread waiting on the GPU.
    }
    \Description{
    Bird's-eye view of the \emph{GPU First} methodology.
    All components on a grid background are provided or generated by the approach. 
    The loader is the entry point for the operating system and responsible to setup the environment on the device.
    The application executable (top right) is produced from the unmodified legacy source code but runs on the GPU.
    A partial \texttt{libc} GPU implementation provides relatively fast device side runtime calls while other library calls are translated into remote procedure calls (RPCs).
    Our RPC scheme will also orchestrate memory movement for arguments and underlying objects, forward the calls to existing system libraries, and return the result to the application thread waiting on the GPU.
    }
    \label{fig:high_level_overview}
\end{figure}

To overcome these challenges, we propose a novel compilation scheme we call ``\emph{GPU First}'' which puts a legacy application on the GPU before any manual porting effort has been started.
Our approach, sketched in \Cref{fig:high_level_overview}, leverages the portability of LLVM's OpenMP offloading to directly compile and run a host application for a GPU, without any source modification.
Instead, the application is compiled for the GPU architecture and started using a provided GPU loader.
Library calls inside the application are either resolved through our partial \texttt{libc} GPU implementation or via automatically generated remote procedure calls (RPCs) to the host.
By adopting the \emph{GPU First} approach, users can seamlessly test and profile their application directly on GPU hardware.
While the performance of full applications running on (current) GPUs is generally not better than CPU execution, it allows developers to easily identify how well existing parallel regions map to the GPU.
Furthermore, it enables rapid testing of code modifications, e.g., data layout transformations or iteration order modifications, on real GPU hardware.
We believe this significant simplification will facilitate the adoption of GPU acceleration in various domains and help developers harness the full potential of modern hardware. 

To evaluate the effectiveness of our approach, we performed experiments on two HPC proxy applications with OpenMP CPU and GPU parallelism, four micro benchmarks with originally GPU only parallelism, as well as three benchmarks from the SPEC OMP 2012 suite featuring hand-optimized OpenMP CPU parallelism.
Our results demonstrate that transparent porting is possible and exploration of GPU performance is feasible for non-experts.
For existing parallel loops, we can closely match the performance of corresponding manually offloaded kernels, achieving up to $14.36\times$ speedup on the GPU compared to the CPU implementation of the HPC proxy application.
This validates our assumptions that the \emph{GPU First} methodology can effectively guide porting efforts, e.g., by identifying parallel regions that require reorganization to achieve good scaling behavior on the GPU, and by allowing fast comparison of different algorithmic and implementation choices.

The main contributions of this paper are summarized here while limitations are explained in detail in \Cref{sec:limitations}.

\setlist[enumerate]{leftmargin=15pt}
\begin{enumerate}
\item A novel compilation scheme that allows to target GPUs for a large set of legacy CPU applications by automatically enabling host-only library calls via a generated RPC interface that translate arguments and mitigate underlying memory for use in dedicated memory environments.
\item A parallelism expansion scheme that allows to map OpenMP parallel directives (and parallel loops) from a single thread block (aka. work group), which is the natural OpenMP offload mapping, to the entire GPU for realistic performance studies.
\item A GPU-optimized partial \texttt{libc} implementation that allows fast execution of runtime calls that do not require operating system support, including a GPU-optimized allocator.
\item An evaluation of the \emph{GPU First} approach on SPEC OMP 2012 benchmarks as well as HPC proxy applications. The former shows the applicability of our scheme to large codes while the latter highlights how parallel loops of the CPU version perform very similar to the manually offloaded GPU kernels. Thus, \emph{GPU First} is well suited to guide porting efforts for legacy applications that already use OpenMP parallelism on the CPU.
\end{enumerate}

The rest of the paper is organized as follows.
In \Cref{sec:background}, we provide background information on OpenMP target offloading and the RPC mechanism.
In \Cref{sec:implementation}, we describe the design and conceptual implementation of our \emph{GPU First} method.
\Cref{sec:limitations} discusses the limitations of our approach and potential directions for future work.
In \Cref{sec:evaluation}, we present our evaluation results to demonstrate the effectiveness of our approach to guide porting efforts for legacy applications.
We will talk about related works in \Cref{sec:related-works} before we conclude with \Cref{sec:conclusion} and insights on the potential impact of our work on the field of parallel computing.
\begin{figure}[b]
\vspace*{-3mm}
\resizebox{.9\linewidth}{!}{
\begin{tikzpicture}[node distance=10mm]

\draw[draw=none,fill=col5!20] (-0.5, 2.35) rectangle (3.5, -3.6);
\draw[draw=none,fill=col6!20] ( 3.5, -1.3) rectangle (7, -3.6);

\node[scale=1,anchor=center,color=gray] (times_txt) at (3.25, -3.40) {compile time / runtime\phantom{p}};
\node[scale=1,anchor=north,align=center,color=gray] (ext_txt) at (3.15, -3.6) {extended LLVM parts};

\node[scale=1.0] (app_source) at (-1.7, 0.0) {\Huge\faIcon[regular]{copy}};
\node[scale=1.0,text width=20mm,anchor=north] (app_source_txt) at ($(app_source.south) + (0.3,0.0)$) {legacy CPU app. source};

\node[scale=1.0] (main_wrapper) at (0.5, 1.5) {\Huge\faIcon[regular]{file-code}};
\node[scale=1.0,anchor=south,yshift=-1mm] (main_wrapper_text) at (main_wrapper.north) {main wrapper};

\node[scale=1.0] (user_wrapper) at (2.5, 1.5) {\Huge\faIcon[regular]{file-code}};
\node[scale=1.0,anchor=south,yshift=-1mm] (user_wrapper_text) at (user_wrapper.north) {user wrapper};

\node[scale=1.3] (compiler) at (1.5, 0.0) {\Huge\faIcon[solid]{tools}};
\node[scale=1.0,text width=3.1cm, anchor=north] (compiler_text) at (compiler.south) {Clang with custom link-time-optimizations};

\node[scale=1.0] (libc) at (1.5, -2.3) {\Huge\faIcon[regular]{file-code}};
\node[scale=1.0,anchor=north] (libc_text) at (libc.south) {partial libc};

\node[scale=1.0] (executable) at (4.3, 0.0) {\Huge\faIcon[regular]{save}};
\node[scale=1.0,anchor=north] (executable_txt) at (executable.south) {exec.};

\node[scale=1.0] (gpu) at (6.1, 0.0) {\Huge\faIcon[regular]{hdd}};
\node[scale=1.0,anchor=north,yshift=-0.8mm] (gpu_txt) at (gpu.south) {GPU};

\node[scale=1.0] (libomptarget) at (4.3, -2.3) {\Huge\faIcon[regular]{file-code}};
\node[scale=1.0,anchor=north] (libomptarget_text) at (libomptarget.south) {offload lib.};

\node[scale=1.0] (rpc) at (6.1, -2.3) {\Huge\faIcon[regular]{hourglass}};
\node[scale=1.0,anchor=north] (rpc_txt) at (rpc.south) {RPC thread};

\draw[line width=2.0pt, ->] ($(app_source.east) + (0.2, 0.0)$) -- ($(compiler.west) + (-0.2, 0)$);
\draw[line width=2.0pt, ->] ($(compiler.east) + (0.2, 0.0)$) -- ($(executable.west) + (-0.2, 0)$);
\draw[line width=2.0pt, ->] ($(executable.east) + (0.2, 0.0)$) -- ($(gpu.west) + (-0.2, 0)$);
\draw[line width=2.0pt, ->] ($(main_wrapper.south) + (0.0, 0.0)$) -- ($(compiler.north west) + (0.0, 0.0)$);
\draw[line width=2.0pt, ->] ($(user_wrapper.south) + (0.0, 0.0)$) -- ($(compiler.north east) + (0.0, 0.0)$);
\draw[line width=2.0pt, ->] ($(libc.north) + (0.0, 0.0)$) -- ($(compiler_text.south) + (0.0, 0.0)$);
\draw[line width=2.0pt, dotted, <->] ($(libomptarget.east) + (0.2,-0.1)$) -- ($(rpc.west) + (-0.2,-0.1)$);
\draw[line width=2.0pt, <->] ($(rpc.north) + (0.0, 0.0)$) -- ($(gpu_txt.south) + (0.0, 0.0)$);
\draw[line width=2.0pt, <->] ($(libomptarget.north) + (0.0, 0.0)$) -- ($(executable_txt.south) + (0.0, 0.0)$);

\end{tikzpicture}
}
\vspace*{-3mm}
\caption{Overview of the compilation and execution path of the direct GPU compilation framework introduced by \citet{DBLP:conf/llvmhpc/TianHPCD22}.
In our work the source of the application still remains unchanged.
The source wrapper and user wrapper files are taken from the original direct GPU compilation paper.
The \texttt{libc} implementation has been extended for this work, and the compiler is now augmented to automatically generate RPC calls and expand source parallelism to the entire GPU device.
The figure was adapted from Figure 1 in \citet{DBLP:conf/iwomp/TianHTCD22}.
The highlighted parts are our contribution.}
\Description{
Overview of the compilation and execution path of the direct GPU compilation framework introduced by \citet{DBLP:conf/llvmhpc/TianHPCD22}.
In our work the source of the application still remains unchanged.
The source wrapper and user wrapper files are taken from the original direct GPU compilation paper.
The \texttt{libc} implementation has been extended for this work, and the compiler is now augmented to automatically generate RPC calls and expand source parallelism to the entire GPU device.
The figure was adapted from Figure 1 in \citet{DBLP:conf/iwomp/TianHTCD22}.
The highlighted parts are our contribution.
}
\label{fig:compilation_execution_path}
\end{figure}
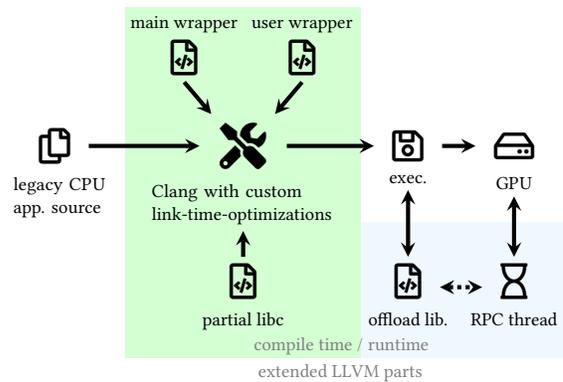

\section{Background}
\label{sec:background}

With the introduction of the \lstinline{target} construct in OpenMP 4.0, it became possible to execute a code region on a target device like a GPU~\cite{DBLP:conf/sc/BertolliAEOSJCS14} or FPGA~\cite{DBLP:conf/iwomp/MayerKP19}.
In this section, we will provide an overview of the LLVM/OpenMP execution model and explain the compilation and execution path of the \emph{GPU First} methodology. 
Additionally, we will introduce the basics of host remote procedure calls (RPCs).

\subsection{OpenMP Execution Model}\label{sec:execution-model}

An OpenMP program begins with a single \emph{initial thread} executing sequentially.
When any thread encounters a \lstinline{parallel} construct, it creates a new team with zero or more additional threads, and each of these threads executes the associated code.

The execution of a \lstinline{target} region is similar to the program start.
A single initial thread executes the device code sequentially.
The \lstinline{parallel} construct works again similar, tough, due to the synchronization requirements of the OpenMP standard, compilers will only use threads of the same thread block / workgroup for the new team.
To utilize the entire device, OpenMP introduced the \lstinline{teams} construct which starts a league of teams each with an independent initial thread.
To distribute work across the league and a team one can use the \lstinline|distribute| and \lstinline|for| constructs, or, alternatively, the global coordinates for manual work-sharing.

%
%

\subsection{Compilation and Execution Path}

Our methodology is based on the compilation and execution path proposed by \citet{DBLP:conf/llvmhpc/TianHPCD22} as part of their direct GPU compilation framework.
Similar to their approach, our methodology compiles the user application for the GPU without modifying the application source code.
However, we augment the compiler in two distinct ways to achieve portability and performance.
First, we automatically generate RPC calls and, second, we expand eligible source parallelism to the entire GPU.
In addition, we enhance their wrapper scripts and partial \texttt{libc} implementation as discussed in \Cref{sec:allocators}.
\Cref{fig:compilation_execution_path}, adapted from \citet{DBLP:conf/llvmhpc/TianHPCD22}, provides an overview of our augmented compilation and execution path.
The main and user wrapper fulfill the same functionality as described by \citet{DBLP:conf/llvmhpc/TianHPCD22}, namely to compile the application code for the GPU architecture, load the environment, e.g, command line options, onto the device and finally transfer control to the user provided \lstinline|main| function on the GPU.
The partial \texttt{libc} implementation is embedded into the application during compile time.
The LLVM/OpenMP offloading runtime, orchestrates the offloading and provides necessary functionality to the RPC host server when it communicates with the GPU threads via ``shared'', in our case, managed, memory.

\subsection{Host Remote Procedural Call}
One of the key challenges in executing a CPU program on a GPU are external functions that are only defined in system or third-party libraries.
We employ remote procedural call (RPC) to utilize the existing host functions during the device execution.
In principle, this allows GPU code to call any function on the host almost as if they were local.
A crucial requirement is the coordination between the host and GPU, that can be achieved through a synchronous, stateless client-server protocol~\cite{DBLP:conf/asplos/SilbersteinFKW13,DBLP:conf/ipps/MikushinLZB14,DBLP:conf/llvmhpc/TianHPCD22}.
In this protocol, the GPU (client) sends requests to the host (server) and waits for the host to acknowledge the completion of the requested function.
Data transfer, e.g., for the arguments and return value, are handled either explicitly \cite{DBLP:conf/llvmhpc/TianHPCD22} or implicitly \cite{DBLP:conf/ipps/MikushinLZB14}.

\section{Design and Implementation}
\label{sec:implementation}

At the core of the \emph{GPU First} methodology is a novel compilation technique that automatically generates a GPU executable from a CPU application with (almost) arbitrary library calls, e.g., to a system or third-party library.
The most complex component of this scheme is the automatic generation of RPCs to transparently execute library functions on the host since they are not available on the GPU.
During compile time, external function calls are replaced by RPC code to perform the call remotely.
When this happens, the call arguments are automatically transferred to the host and, in case of pointers, the underlying memory is usually migrated as well.
While not infallible, our proof-of-concept scheme automatically handles the most common cases.
Limitations are discussed in more detail in \Cref{sec:limitations}.
By minimizing, often even eliminating, the manual work required to run applications on the GPU, our approach aims to facilitate the adoption of GPU acceleration and help non-experts leverage the potential of modern hardware.

\subsection{GPU Code Generation and Loading}

In contrast to classical offloading we directly target the GPU with the entire application code.
For this task we utilize an extended version of the direct GPU compilation framework by \citet{DBLP:conf/llvmhpc/TianHPCD22}.
In a nutshell, the application code is transparently enclosed in a \lstinline{#pragma omp begin/end declare target device_type(nohost)}\\ scope to utilize existing OpenMP offloading support in LLVM/ Clang to generate GPU code.
In the final executable, the application \lstinline{main} function is invoked via the OpenMP offload mechanism after command line arguments have been mapped to the device and the host RPC server has been set up.
For more information on the code generation and loading we refer to the original paper~\cite{DBLP:conf/llvmhpc/TianHPCD22}.

\subsection{Generating Remote Procedural Calls}
\label{sec:rpc-and-memory}

To identify and efficiently replace library calls in the application by RPCs we provide a dedicated link time optimization (LTO) pass.
The benefit over per translation unit reasoning is the complete world view which includes all the functions defined by the application as well as the call sites and contexts in which library functions, or simply non-defined functions, are called.
It is important to note that compared to existing LLVM-IR passes our RPC generation emits host code while the device code is analyzed and transformed.

\begin{figure}[p]
\lstset{ numbers = right,
         numbersep = 5pt, 
         numberstyle = {\color{gray}},
         stepnumber = 4,
        basicstyle=\footnotesize\lst@ifdisplaystyle\footnotesize\linespread{0.94}\fi\ttfamily,
        }
         
\ContinueLineNumber

\begin{minipage}{\linewidth}
\vspace{2mm}
\begin{lstlisting}[numberfirstline=true, stepnumber=5, frame=lines]
struct S { int a, b; float f; };
void use(struct S* s, int r, int i) { ... }
void example(struct S s, int *p) {
  int i;
  int r = fscanf(fd, "%f %i %i", &s.f, S.a ? &i : &S.b, p);
  use(&s, r, i);
}
\end{lstlisting}
\vspace{-3mm}
\subcaption{
Variadic function call that will read and write memory passed in via pointers.
The underlying memory is located on the stack (\lstinline|i| and \lstinline|s|), inside of a larger object (\lstinline|s.b|, \lstinline|s.f| inside \lstinline|s|), or part of a statically unknown object (via \lstinline|p|).
}
\label{fig:rpc_gen_src}
\vspace{3mm}
\end{minipage}

\begin{minipage}{\linewidth}
\begin{lstlisting}[frame=lines]
int __fscanf_ip_fp_ip(RPCInfo &RI) {
  auto fd =       (FILE*)RI.getArg(0);
  auto fm = (const char*)RI.getArg(1);
  auto f2 =      (float*)RI.getArg(2);
  auto i3 =        (int*)RI.getArg(3);
  auto i4 =        (int*)RI.getArg(4);
  return fscanf(fd, fm, f2, i3, i4);
}
\end{lstlisting}
 \vspace{-3mm}
\subcaption{Host code generated during compilation. The the variadic function call is embedded in a non-variadic function named based on the call argument types.}
\label{fig:rpc_gen_host}
\vspace{3mm}
\end{minipage}

\begin{minipage}{\linewidth}
\begin{lstlisting}[frame=lines]
int __fscanf_ip_fp_ip(RPCArgInfo &RAI) {
  RPCInfo RI;
  RI.setCallee(/* compile time */ enum(__fscanf_ip_fp_ip));
  RI.setArgInfo(&RAI);
  return RI.issueBlockingCall();
}
  
void example(struct S s, int *p) {
  int i;
  RPCArgInfo CallSiteRAI(/* num args */ 5);
  
  // Opaque value, treated as byte sequence:
  CallSiteRAI.addValArg(fd);
  
  // Statically identified objects:
  CallSiteRAI.addRefArg("%f %i %i", read,       (*@ \hfill{} @*) \
              sizeof("%f %i %i"), /* offset */ 0);
  CallSiteRAI.addRefArg(&s.f, readwrite,    (*@ \hfill{} @*) \
              sizeof(s), offsetof(struct S, f));
  if ((s.a ? &i : &s.b) == &i) (*@ \label{fig:rpc_gen_device_compile_conditional_begin} @*)
    CallSiteRAI.addRefArg(&i, write, sizeof(i), 0);
  else /* ((s.a ? &i : &s.b) == &s.b) */
    CallSiteRAI.addRefArg(&s.b, readwrite,     (*@ \hfill{} @*) \
              sizeof(s), offsetof(struct S, b)); (*@ \label{fig:rpc_gen_device_compile_conditional_end} @*)
  
  // Statically unknown object requires dynamic lookup:
  int p_offset, p_size;
  if (_FindObj(p, &p_offset, &p_size, PotentialObjs))
    CallSiteRAI.addRefArg(p, readwrite, p_size, p_offset);
  else
    CallSiteRAI.addValArg(p);
                    
  int r = __fscanf_ip_fp_ip(CallSiteRAI);
  use(&s, r, i);
}
\end{lstlisting}
\vspace{-3mm}
\subcaption{
The figure displays the device code that is generated for the example shown in \Cref{fig:rpc_gen_src}, where the argument information at the call site is both statically and dynamically embedded into the \lstinline|RPCArgInfo| object.
This information, along with the call site independent information, guides the RPC system to determine which function needs to be called on the host, what memory needs to be copied, and how arguments should be translated.
}
\label{fig:rpc_gen_device}
\vspace{-1mm}
\end{minipage}

\caption{
Example illustrating the host and device code generated during compile time to substitute a variadic library function call site in the user code with RPC and memory management code.
}
\Description{
Example illustrating the host and device code generated during compile time to substitute a variadic library function call site in the user code with RPC and memory management code.
}
\label{fig:rpc_gen}
\end{figure}

In the following we will walk through the compile time generation of RPC calls using \Cref{fig:rpc_gen} as a guide.
The top part, \Cref{fig:rpc_gen_src}, shows a manufactured example of user code that exhibits most complexities and optimization opportunities we support.
The variadic library call to \lstinline{fscanf} is replaced with an RPC call on the device, and a wrapper that is invoked by the RPC server thread on the host.
The latter, shown in \Cref{fig:rpc_gen_host}, unpacks the arguments passed from the device and performs the original call on the host.
Arguments are stored in an opaque fashion inside the \lstinline{RPCInfo} object that is used for communication and (bit-)casted to their respective type.
For variadic callees, the host wrapper function name uses the variadic argument types to provide different entry points for variadic call sites that do not agree on the number of arguments or their type.
Said differently, for variadic function calls we effectively generate a non-variadic landing-pad on the host for each combination of call site argument types we encounter.
For non-variadic function calls there is a unique combination of argument types paired corresponding to a single host function.

The device code that replaces the original library call, shown in \Cref{fig:rpc_gen_device}, is divided in call site specific code and call site independent code.
The latter is the \lstinline{__fscanf_ip_fp_ip} function that issues the RPC call and waits for the result.
Information about arguments is provided in the \lstinline|RPCArgInfo| object.
The callee is identified through a compile time generated enum value representing the function issuing the RPC.
The call site specific code records information about the arguments to orchestrate data transfer of memory, as it is potentially required for the library call.
Dedicated call site information allows for more efficient code if call sites disagree what (dynamic) argument value is passed to a library function.
In general, there are three different kinds of arguments.
The simplest are value arguments, which include integers, floating point values, as well as pointers to opaque types.
The first call argument in our example (\lstinline|fd|) is of the latter type; namely a \lstinline|FILE *|. 
Since we do not know what a \lstinline|FILE| object looks like, we assume pointers to them are only accessible on the host and the associated memory is never moved to the device.
Said differently, we assume the pointer is pointing to host memory already and consequently does not need translation for the RPC.
Thus, the value of \lstinline|fd| will be exactly the same on the host as it is on the device.
The second kind of arguments are pointers to statically identified objects by applying inter-procedural analysis built on top of the LLVM's Attributor framework.
Those objects can (conceptually) reside in stack, global, or constant memory.
The next three call arguments in our example (\lstinline{"
The format string is a known compile time constant but we still need to make it available on the host such that \lstinline|fscanf| can read it.
To this end, we register the pointer together with the size of the underlying object, here \lstinline|sizeof("%f %i %i")|, and the offset of the pointer into the object, here \lstinline|0|.
Since the object is constant, we mark it as read only which ensure the memory is only copied to the host and not back.
After the memory has been transferred, we will use a pointer with the same offset into the host version of the object for the host call site.
The second argument is pointing to a statically identified object is \lstinline|&s.f|.
The handling is similar as before but the offset into the object is not trivial this time.
Further, the memory might be read and written by \lstinline|fscanf| which means we need to copy the object to the host and back.
For the next call site argument we can not statically determine a unique underlying object but we can enumerate all possibilities statically and they are all statically identified objects.
Since offset and size are different we need to generate code that identifies the object at runtime based on the pointer value, as shown in lines \ref{fig:rpc_gen_device_compile_conditional_begin}-\ref{fig:rpc_gen_device_compile_conditional_end} in \Cref{fig:rpc_gen_device}.
Note that \lstinline|&i| is marked write only since the underlying object is known to have an unspecified value on the device at the time of the call.
The last category of arguments are pointers for which we can not statically enumerate all potential underlying objects.
This might be because the pointer is a result of a \lstinline|malloc|-like call, which can be executed multiple times in a loop resulting in different objects, a stack allocation inside a potentially recursive functions, which also allows for multiple runtime instantiations, or a pointer with an unknown origin.
In all cases we will attempt to identify the underlying object at runtime in order to determine the size and offset of the pointer.
To accomplish this, we rely on our allocator, which we use to implement \lstinline|malloc|-like calls, to maintain a record of allocated objects, as discussed in more detail as part of \cref{sec:allocators}.
In case we are unable to determine the object, we will treat the pointer as a value assuming that the it is not accessed or already points to host memory.

\subsection{Multi-Team Execution and Kernel Split}

The original direct GPU offloading work was unable to predict performance of offloaded versions accurately due because the offloaded host version was run with a single team (or thread block)~\cite{DBLP:conf/llvmhpc/TianHPCD22}.
Workload distribution in a single team, either explicitly coded with \lstinline|omp_get_thread_num| and \lstinline|omp_get_num_threads| or automatically applied via \lstinline|omp for|, will only utilize threads of the team, which is insufficient for realistic scaling studies on a GPU.
However, the workload of many parallel regions can be executed by multiple teams without violating the program semantics.
To address this, we implemented a compiler transformation that can identify and convert amendable parallel regions into kernels executed by multiple teams (or thread blocks).
For each such parallel region, the schedule strategy of each automatic work-sharing construct (\lstinline|omp for|) is changed to distribute the work across threads in all teams.
This is similar as rewriting an \lstinline|omp for| into an \lstinline|omp distribute (parallel) for|.
For manual worksharing we replace the query calls for thread id and number of threads with versions that take the threads of all blocks into account.
In both cases the rewrite ensures the workload is distributed among all the teams and all the threads in each team.
Similarly, \lstinline|omp barrier| constructs need to be replaced by a version that synchronizes across all teams (or thread blocks).
While that is not allowed by the OpenMP standard, modern GPUs provide means to achieve this in practice, e.g., via global atomic counters.

For the sequential part of the original application we still utilize a single team as there is no need for more threads and any additional team would require special handling to guard against side effects caused by its initial thread.
Whenever the initial thread encounters a parallel region that has been converted to a multi-team kernel.
We will issue an RPC call to launch the kernel from the host with the same arguments the parallel region would have been given.
The basic idea is to replace an \lstinline|omp parallel| with a \lstinline|omp teams|, followed by an \lstinline|omp parallel|, and appropriate changes to the parallel region, especially the worksharing parts and synchronization.

Figure \ref{fig:parallel-kernel} illustrates the concepts of multi-team execution and kernel splitting.
On the left-hand side, a single team executes the main kernel, consisting of one main thread and four worker threads.
Initially, only the main thread runs while the worker threads are waiting in a state machine, as described in \cite{DBLP:conf/cgo/HuberCGTDDCD22}.
When a parallel region is encountered, the worker threads start executing, and the main thread waits for the region to complete.
Once the parallel region is finished, the main thread resumes the serial part of the program, while the worker threads wait for their next task.
On the right-hand side, multi-team execution is used.
The main kernel has only one team with one thread.
When a parallel region is encountered, a host RPC call is issued with all the necessary information to launch a parallel kernel with multiple teams.
The main thread waits for the host RPC to finish before proceeding.
In this example, four teams are launched, each with four threads.
These teams are not taken as separate entities; instead, they are bulked together as one large team, ensuring that all the threads have continuous thread IDs, rather than starting from 0 in each team.
Once the parallel kernel is completed, the host RPC call is finished, and the main thread in the main kernel proceeds with the execution of the serial part of the program.

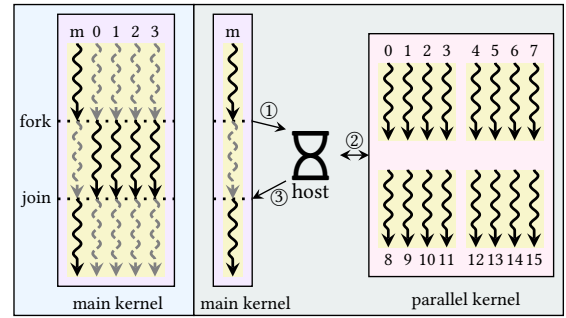
\begin{figure}[htb]
\vspace{-2mm}
\resizebox{0.9\linewidth}{!}{
\begin{tikzpicture}[node distance=10mm]

\def\mainkernelleftcornerx{-2mm}

\draw[fill=col6!25] (-6.5mm, 4mm) rectangle (12mm, -28mm);
\draw[fill=col13!8] (12mm, 4mm) rectangle (50mm, -28mm);

\draw[fill=col1!20] (-2mm, 3mm) rectangle (10mm, -25mm);
\draw[fill=col2!20,draw=none] (-1mm,0mm) rectangle (9mm, -24mm);
\draw[fill=col1!20] (14mm, 3mm) rectangle (18mm, -25mm);
\draw[fill=col2!20,draw=none] (15mm,0mm) rectangle (17mm, -24mm);
\node[scale=0.6,anchor=south east] at (9.5mm, -28mm) {main kernel};

\node[scale=0.6,anchor=south] at (0mm, 0mm) {m};
\draw[thread] (0mm, 0mm) -- (0mm, -8mm);
\foreach \c [evaluate=\c as \tid using int(\c/2-1)] in {2, 4, 6, 8} {
\node[scale=0.6,anchor=south] (n) at (\c mm, 0mm) {\tid};
\draw[workerthread] (n.south) -- (\c mm, -8mm);
}

\draw[thick, dotted] (-2mm, -8mm) -- (10mm, -8mm);
\draw[thick, dotted] (14mm, -8mm) -- (18mm, -8mm);

\node[scale=0.6,anchor=east] at (-2mm, -8mm) {fork};

\draw[workerthread] (0mm, -8mm) -- (0mm, -16mm);
\foreach \c in {2, 4, 6, 8} {
\draw[thread] (\c mm, -8mm) -- (\c mm, -16mm);
}

\draw[thick, dotted] (-2mm, -16mm) -- (10mm, -16mm);
\draw[thick, dotted] (14mm, -16mm) -- (18mm, -16mm);

\node[scale=0.6,anchor=east] at (-2mm, -16mm) {join};

\draw[thread] (0mm, -16mm) -- (0mm, -24mm);
\foreach \c in {2, 4, 6, 8} {
\draw[workerthread] (\c mm, -16mm) -- (\c mm, -24mm);
}


\node[scale=0.6,anchor=south] at (16mm, 0mm) {m};
\draw[thread] (16mm, 0mm) -- (16mm, -8mm);
\draw[workerthread] (16mm, -8mm) -- (16mm, -16mm);
\draw[thread] (16mm, -16mm) -- (16mm, -24mm);

\node[scale=0.6,anchor=south west] at (12mm, -28mm) {main kernel};

\node[] (host) at (24mm, -11.5mm) {\huge\faIcon[regular]{hourglass}};
\node[scale=0.7] at (host.south) {host};

\draw[->] ($(16mm, -8mm) + (2mm, 0mm)$) -- node [above,scale=0.6] {\textcircled{\raisebox{-.7pt}{1}}} ($(host.north west) + (0.5mm, -1mm)$);
\draw[->] ($(host.south west) + (0.5mm, 1mm)$) -- node [below,scale=0.6,pos=0.2] {\textcircled{\raisebox{-.7pt}{3}}} ($(16mm, -16mm) + (2mm, 0mm)$);

\draw[<->] (host.east) -- node [above,scale=0.6] {\textcircled{\raisebox{-.7pt}{2}}} ($(host.east) + (3mm, 0mm)$);

\draw[fill=col4!20] (30mm,1mm) rectangle (49mm, -24mm);
\node[scale=0.6,anchor=south] at (40mm, -28mm) {parallel kernel};

\draw[fill=col2!20,draw=none] (31mm, -2mm) rectangle (39mm, -10mm);
\foreach \c [evaluate=\c as \tid using int(\c/2)] in {0, 2, 4, 6} {
\node[scale=0.6,anchor=south] (n) at ($(32 mm, -2mm) + (\c mm, 0mm)$) {\tid};
\draw[thread] (n.south) -- ($(32 mm, -10mm) + (\c mm, 0mm)$);
}

\draw[fill=col2!20,draw=none] (40mm, -2mm) rectangle (48mm, -10mm);
\foreach \c [evaluate=\c as \tid using int(\c/2 + 4)] in {0, 2, 4, 6} {
\node[scale=0.6,anchor=south] (n) at ($(41 mm, -2mm) + (\c mm, 0mm)$) {\tid};
\draw[thread] (n.south) -- ($(41 mm, -10mm) + (\c mm, 0mm)$);
}

\draw[fill=col2!20,draw=none] (31mm, -13mm) rectangle (39mm, -21mm);
\foreach \c [evaluate=\c as \tid using int(\c/2+8)] in {0, 2, 4, 6} {
\node[scale=0.6,anchor=north] (n) at ($(32 mm, -21mm) + (\c mm, 0mm)$) {\tid};
\draw[thread] ($(32 mm, -13mm) + (\c mm, 0mm)$) -- (n.north);
}

\draw[fill=col2!20,draw=none] (40mm, -13mm) rectangle (48mm, -21mm);
\foreach \c [evaluate=\c as \tid using int(\c/2+12)] in {0, 2, 4, 6} {
\node[scale=0.6,anchor=north] (n) at ($(41mm, -21mm) + (\c mm, 0mm)$) {\tid};
\draw[thread] ($(41mm, -13mm) + (\c mm, 0mm)$) -- (n.north);
}

\end{tikzpicture}
}
\vspace{-2mm}
\caption{Illustration of multi-team execution and kernel splitting.
The left side shows single team execution with one main thread and four worker threads.
The right side shows multi-team execution with one main thread in the main kernel and four teams of four threads each in the parallel kernel.
A host RPC call (\textcircled{\raisebox{-.7pt}{1}} and \textcircled{\raisebox{-.7pt}{3}}) is used to launch the parallel kernel (\textcircled{\raisebox{-.7pt}{2}}) with multiple teams.}
\Description{
Illustration of multi-team execution and kernel splitting.
The left side shows single team execution with one main thread and four worker threads.
The right side shows multi-team execution with one main thread in the main kernel and four teams of four threads each in the parallel kernel.
A host RPC call (\textcircled{\raisebox{-.7pt}{1}} and \textcircled{\raisebox{-.7pt}{3}}) is used to launch the parallel kernel (\textcircled{\raisebox{-.7pt}{2}}) with multiple teams.
}
\label{fig:parallel-kernel}
\end{figure}

\subsection{Heap Allocators and Allocation Tracking}
\label{sec:allocators}

We have extended the partial GPU \lstinline{libc} implementation by \citet{DBLP:conf/llvmhpc/TianHPCD22} to provide more functions that can run natively on GPUs without the need for RPCs.
The extensions were guided by benchmarks and include functions such as \lstinline{strtod}, \lstinline{rand}, and \lstinline|realloc|.
Additionally, we provide our own implementations of \lstinline{malloc} that the user can choose from via the compile time flag:\\
\texttt{-fopenmp-target-allocator=\{generic,balanced[N,M]\}}

There are two main reasons for configurable custom allocators.
Firstly, the support for heap allocation on GPUs varies among vendors, and any one implementation can not handle all situations optimally.
To improve the end-to-end execution time the user needs to choose an allocator implementation based on their particular use case.
As an example, one of our benchmarks has massively parallel heap allocations and deallocations at the beginning, and respectively end, of a parallel region.
Serialization can cause significant delays.
Other benchmarks require large amounts of heap memory but only use the initial thread to allocate it.
This means the heap space can not be divided into exclusive parts (per thread and/or team).
Secondly, we need to track allocated memory regions to support the runtime lookup for underlying objects in case they can not be determined at compile time, as mentioned in \Cref{sec:rpc-and-memory}.

\begin{figure}[htb]
\vspace{-1mm}
\resizebox{.9\linewidth}{!}{
\newcommand{\WAEntry}[4]{%
\draw[fill=#3] 
      ($(#1, 4) + (0,0)$) rectangle ($(#1, 0)+ (4,0)$);
\draw[fill=#3, #4] 
      ($(#1, 4) + (4,0)$) rectangle ($(#2, 0)+ (0,0)$);
\draw[line width=2.5pt] ($(#1, 4) + (0,0)$) -- ($(#1, 0) + (0, 0)$);
\draw[dashed] ($(#1, 4) + (2,0)$) -- ($(#1, 0)+ (2,0)$);
\draw[] ($(#1, 4) + (4,0)$) -- ($(#1, 0)+ (4,0)$);
\draw[line width=2.5pt] ($(#2, 4) + (0,0)$) -- ($(#2, 0)+ (0,0)$);
}
\newcommand{\WALinkFwd}[3]{%
\draw[line width=13pt,white] ($(#1, 5) + (3, 0)$) -- ($(#2, 5) + (-0.2, 0)$);
\draw[line width=5pt,->] ($(#1, 2) + (3,0)$) -- ($(#1, 5) + (3, 0)$) -- ($(#2, 5) + (0, 0)$) -- ($(#2, 4) + (0, 0.1)$);
}
\newcommand{\WALink}[3]{%
\WALinkFwd{#1}{#2}{#3}
\draw[line width=13pt,white] ($(#2, 5.5) + (1, #3)$) -- ($(#1, 5.5) + (0.2, #3)$);
\draw[line width=5pt,->] ($(#2, 2) + (1,0)$) -- ($(#2, 5.5) + (1, #3)$) -- ($(#1, 5.5) + (0, #3)$) -- ($(#1, 4) + (0, 0.1)$);
}

\begin{tikzpicture}[node distance=10mm]

\draw[] (0, 4) -- (40, 4);
\draw[] (0, 0) -- (40, 0);
\WAEntry{4}{14}{col1!30}{postaction={pattern color=gray, pattern=crosshatch}}
\WAEntry{14}{23}{col2!30}{postaction={pattern color=gray, pattern=crosshatch}}
\WAEntry{24}{38}{col3!30}{postaction={pattern color=gray, pattern=crosshatch}}
\WALink{14}{24}{1}
\WALink{4}{14}{0}
\WALinkFwd{24}{38}{0}
\node[scale=1.4,anchor=north, rotate=90] at (23, 2) {\Huge\textbf{padding}};
\draw[line width=4.5, <-|,dashed] (4, 0) -- (4, -1) -- (6, -1) node[scale=2.0, anchor=west, yshift=1.5pt] {\Huge{bottom}};
\draw[line width=4.5, <-|,dashed] (38, 0) -- (38, -1) -- (36, -1) node[scale=2.0, anchor=east, yshift=-1.5pt] {\Huge{top}};

\tikzset{shift={(0,-9)}}

\draw[] (0, 4) -- (40, 4);
\draw[] (0, 0) -- (40, 0);
\WAEntry{4}{14}{col1!30}{postaction={pattern color=gray, pattern=crosshatch}}
\WAEntry{14}{23}{col2!30}{}
\WAEntry{24}{38}{col3!30}{postaction={pattern color=gray, pattern=crosshatch}}
\WALink{14}{24}{1}
\WALink{4}{14}{0}
\WALinkFwd{24}{38}{0}
\node[scale=1.4,anchor=north, rotate=90] at (23, 2) {\Huge\textbf{padding}};
\draw[line width=4.5, <-|,dashed] (4, 0) -- (4, -1) -- (6, -1) node[scale=2.0, anchor=west, yshift=1.5pt] {\Huge{bottom}};
\draw[line width=4.5, <-|,dashed] (38, 0) -- (38, -1) -- (36, -1) node[scale=2.0, anchor=east, yshift=-1.5pt] {\Huge{top}};

\tikzset{shift={(0,-9)}}

\draw[] (0, 4) -- (40, 4);
\draw[] (0, 0) -- (40, 0);
\WAEntry{4}{14}{col1!30}{postaction={pattern color=gray, pattern=crosshatch}}
\WALinkFwd{4}{14}{0}

\draw[line width=4.5, <-|,dashed] (4, 0) -- (4, -1) -- (6, -1) node[scale=2.0, anchor=west, yshift=1.5pt] {\Huge{bottom}};
\draw[line width=4.5, <-|,dashed] (14, 0) -- (14, -1) -- (16, -1) node[scale=2.0, anchor=west, yshift=-1.5pt] {\Huge{top}};

\tikzset{shift={(0,-3)}}

\draw[line width=4.5, <-|] (0,0) -- (0.5, 0) node[scale=2.0, anchor=north west] {\Huge{previous heap }} -- (4,0);
\draw[line width=4.5, |->] (14, 0) -- (38,0) node[scale=2.0, anchor=north east] {\Huge{free heap space, then next heap}} -- (40,0);

\end{tikzpicture}
}
\vspace{-2mm}
\caption{Visualization of one chunk in the balanced allocator.
Top: Encoding of three user allocated entries all currently in use, as indicated by the missing grid pattern on the user data.
Middle: Situation after the second entry has been deallocated.
The encoding is not changed to speed up deallocation.
Allocation remains fast as long as sufficient heap space is available to add an entry on top.
Bottom: Situation after the former top entry has been deallocated.
As long as the top entry is unused the space is reclaimed, which makes the scheme especially suitable for balanced allocations and deallocations.
}
\Description{
Visualization of one chunk in the balanced allocator.
Top: Encoding of three user allocated entries all currently in use, as indicated by the missing grid pattern on the user data.
Middle: Situation after the second entry has been deallocated.
The encoding is not changed to speed up deallocation.
Allocation remains fast as long as sufficient heap space is available to add an entry on top.
Bottom: Situation after the former top entry has been deallocated.
As long as the top entry is unused the space is reclaimed, which makes the scheme especially suitable for balanced allocations and deallocations.
}
\label{fig:warp_allocator}
\vspace{-2mm}
\end{figure}
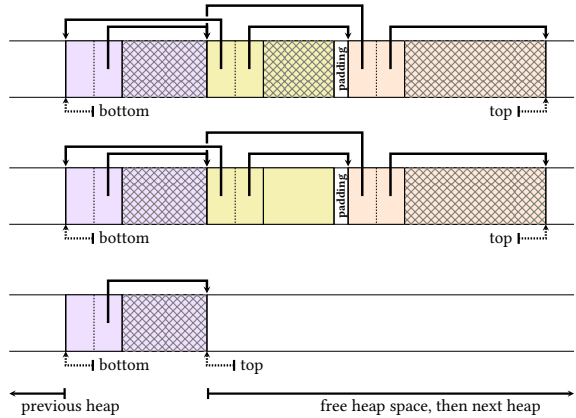

We provide two types of allocators: a single-thread generic allocator and a ``balanced'' allocator.
The single-thread generic allocator tracks all allocations in two linked lists: an allocation list and a free list.
Each thread can use the entire heap space if necessary, but access to the lists has to be mutually exclusive, which can become a performance bottleneck for applications that allocate heap memory concurrently.
The balanced allocator is designed to mitigate this limitation.
It divides the heap space into $N\times{}M$ chunks, with each thread calculating its chunk based on its thread and team id module $N$ and $M$, respectively. 
We use a lock per chunk to ensure consistency, but different chunks are independent.
Since the heap memory each thread can allocate is relatively small, it is possible to run out of memory in only a specific chunk while others are mostly empty.
As it is common to allocate large heap areas in the serial execution part of a program, the first chunk of the $N$ is larger than the rest (with a configurable ratio).
Since the initial thread is always the first thread of a warp / wavefront, it has consequently more heap available when it executes the sequential program parts.

The balanced allocator differs from the generic allocator in that it embeds indices/pointers into the allocation metadata rather than using explicit linked lists.
\Cref{fig:malloc} illustrates this concept for a single chunk.
The top row displays three allocations that are currently in use, indicated by the grid pattern over the user data.
When the middle entry is deallocated, it is marked as such but the now-free memory is kept in place and not referenced elsewhere.
To reuse previously deallocated memory regions that have not been reclaimed, we need to traverse the list until a suitable entry is found, which can be costly in practice.
Thus, we avoid reusing allocations until we exhaust the heap space in this chunk.
We reclaim the top allocation by moving the watermark pointer to the end of the previous entry whenever the top allocation is no longer in use.
The newly formed top may also be reclaimed if it was previously deallocated.
The third row of the figure illustrates the situation after the top entry was deallocated and the two top entries~have~been~reclaimed.

While this scheme is not a replacement for a more generic allocator, it is well-suited for applications with balanced allocations and deallocations in terms of their lifetime since we can reclaim memory with minimal overhead during the allocation and deallocation calls.
In the worst case, interleaved allocations with different lifetimes cause holes and costly linear traversals as soon as the heap space runs out.
However, if we do not run out of heap space, there is likely only minor performance degradation due to fragmentation.

\section{Limitations and Future Works}
\label{sec:limitations}

\emph{GPU First} is a proof-of-concept implementation that showcases how advancements in GPUs, coupled with modern compiler technology, can simplify the approach to GPU programming significantly.
The conventional restrictions, such as the absence of recursion, lack of atomic accesses, and so on, that have led to the current design of offload languages should be reconsidered, and alternative methods, such as \emph{GPU First}, should be explored.
Despite its potential, there are various technical challenges that \emph{GPU First} needs to overcome, which we will discuss briefly.

\subsection{Multiple Levels of Indirection}

In \Cref{sec:rpc-and-memory} we described that we move underlying objects to and from the host when a pointer to them is used in an RPC call.
This allows the library function to access the object, e.g., to write the result from a file I/O operation into that memory.
However, we do not yet try to move more than one level of memory which prevents the host function from accessing objects through indirection.
As an example, a host function might be passed a \lstinline|int**| and we would migrate the object the outer pointer points to automatically.
Thus, when the initial pointer (after translation to the host value) is dereferences, the migrated memory is accessed.
If, however, the resulting \lstinline|int*| is accessed by the host, the value will likely point to device memory.
While it is possible to move and update pointers for multiple levels, the precision and efficiency of the approach will depend on the availability of domain knowledge about accesses.
Annotated library headers, as generated by the ``HTO''~\cite{poster/MosesHTO}, would likely make this feasible in practice.
A system with unified shared would not encounter this problem at all.

\subsection{Reverse Offloading of Code}

In our prototype we only execute host landing-pad functions generated during compilation.
However, we plan to extend this capability in the future to allow for the execution of other host code, such as when a function pointer is passed to the host via an RPC or when an object method is invoked as part of an RPC.
The first step would be to generate potentially host executed code for the host as well, which could be as simple as generating all code for the host and the GPU.
In the second step we need to translate function pointers from the device to the host value when objects are moved from the device to the host, or alternatively when a fault is caused while trying to execute code through a device function pointer.
If objects or function pointers are created on the host as part of an RPC the revere procedure is required.
This shortcoming is less severe in legacy C code but shows when C++ objects are created on the device and used on the host as their virtual table contains both, an additional level of indirection and pointers to device-only code.
That said, C++ objects that are only used on the device are already supported, including virtual function calls and other inheritance-related features.

While the above limits the applicability of the \emph{GPU First} methodology, there are related opportunities to improve performance in the presence of code regions that should be executed on the host.
So far, only single library calls are issued on the host, however, entire code regions in the original application could benefit from execution on the host.
The applicability limitations discussed so far notwithstanding, we could outline the region and treat the code as if it was originally in an external library.
As such, our RPC generation would take care of the call and (single-level) memory movement.

\subsection{Multi-Team Execution with Communication}

Another limitation of our work is that we only rewrite certain parts of the code to support multi-team execution in our prototype.
For example, we change the work-sharing schedule and make sure the user observed thread Ids are continues across the threads in the different teams (ref.~\Cref{fig:parallel-kernel}).
While we do not yet rewrite inter-thread communication, such as \lstinline{reduction} clauses, most common cases could be handled through additional engineering effort.


\subsection{Single-Threaded RPC Handling}

Our prototype features single-threaded RPC handling which can, in parallel regions, significantly impact performance.
However, since multi-threaded RPC schemes can be implemented, this is not a conceptual limitation and will also not influence our benchmarks that do not issue RPC calls from inside parallel regions.

\section{Evaluation}
\label{sec:evaluation}


To evaluate the performance of our approach, we used a system comprising an NVIDIA A100 Tensor Core GPU (40GB) with AMD EPYC 7532 processors (32 cores and hyper-threading disabled) and 256 GB DDR4 RAM.
We performed all experiments using CUDA 11.8.0 and compiled all benchmarks using the \lstinline{-O3} optimization flag.
Our prototype version is based on LLVM trunk (\raisebox{-1.5pt}{\faIcon[regular]{git}} \lstinline{50f1476a}).

\subsection{Allocator Performance}

We believe an application should pick an allocator based on their specific needs.
Some of the evaluated SPEC OMP benchmarks concurrently allocate many memory regions at the beginning of a parallel, and deallocate them at the end of the parallel region again.
To best support this scheme we introduced the balanced allocator described in \Cref{sec:allocators}.
In \Cref{fig:malloc} shows how it performs on a synthetic benchmark in which all threads in all teams allocate memory at the beginning of the kernel, use it briefly, and then deallocate it again.
This design is an exaggeration of the SPEC OMP benchmarks to stress test  allocators.
On our test platform the domain specific balanced allocator is between $3.3\times$ (1 thread, 1 team) and $30\times$ (32 threads, 256 teams) faster than the default NVIDIA provided \lstinline|malloc|.

\begin{figure}[hbt]
    \centering
    \resizebox{.9\linewidth}{!}{
      \begin{tikzpicture}
\begin{axis}[%
    legend style={
      at={(0, 1.3)},
      anchor=north west,
      draw=none,
      fill=none,
    },
      legend cell align={left},
scatter/classes={%
    w1={mark=o,draw=col1,ultra thick,scale=1.6},
    w32={mark=o,draw=col3,ultra thick,scale=1.6},
    w64={mark=o,draw=col5,ultra thick,scale=1.6},
    w128={mark=o,draw=col6,ultra thick,scale=1.6},
    w256={mark=o,draw=col7,ultra thick,scale=1.6},
    c1={mark=x,draw=col1,ultra thick,scale=1.5},
    c32={mark=x,draw=col3,ultra thick,scale=1.5},
    c64={mark=x,draw=col5,ultra thick,scale=1.5},
    c128={mark=x,draw=col6,ultra thick,scale=1.5},
    c256={mark=x,draw=col7,ultra thick,scale=1.5}
    },
    ymode = log,
xtick      ={1,  2,  3,   4,   5},
xticklabels={1, 32, 64, 128, 256},
xlabel = {number of threads per team},
xtick pos=bottom,
    xtick style={line width=20pt,yshift=-2pt},
    ylabel = {execution time in seconds},
    ylabel style = {align=center,at={(-0.12,0.55)},text width=4cm},
]
    
\addlegendimage{only marks,mark=o,ultra thick,scale=1.6}
\addlegendentry{Balanced Allocator (32 thread slots, 16 team slots)}
\addlegendimage{only marks,mark=x,ultra thick,scale=1.5}
\addlegendentry{NVIDIA malloc}

\addplot[scatter,only marks,%
    scatter src=explicit symbolic]%
table[meta=label] {
x y label
0.8 0.000005943 w1 
1.8 0.000008355 w1
2.8 0.000011715 w1
3.8 0.000019677 w1
4.8 0.000040061 w1
0.9 0.000008448 w32
1.9 0.000018851 w32
2.9 0.000027302 w32
3.9 0.000045264 w32
4.9 0.000086458 w32 
1.0 0.000013222 w64
2.0 0.000027923 w64
3.0 0.000046509 w64
4.0 0.000085261 w64
5.0 0.000232615 w64 
1.1 0.000023645 w128
2.1 0.000044717 w128
3.1 0.000086179 w128
4.1 0.000234955 w128
5.1 0.000865904 w128 
1.2 0.000023645 w256
2.2 0.000044717 w256
3.2 0.000086179 w256
4.2 0.000234955 w256
5.2 0.000865904 w256
0.8 0.000020083 c1
1.8 0.000086080 c1
2.8 0.000105703 c1
3.8 0.000154804 c1
4.8 0.000213147 c1
0.9 0.000037354 c32
1.9 0.000361153 c32
2.9 0.000542578 c32
3.9 0.000923219 c32
4.9 0.001336460 c32
1.0 0.000056922 c64
2.0 0.000553829 c64
3.0 0.000865383 c64
4.0 0.001313250 c64
5.0 0.001959345 c64
1.1 0.000093300 c128
2.1 0.000877581 c128
3.1 0.001320811 c128
4.1 0.002102296 c128
5.1 0.003548110 c128
1.2 0.000165137 c256
2.2 0.001336754 c256
3.2 0.002075279 c256
4.2 0.003669188 c256
5.2 0.007008149 c256
};
\end{axis}

\begin{axis}[%
    legend style={
      at={(0.5, 1.08)},
      anchor=center,
      draw=none,
    },
    legend columns=5,
scatter/classes={%
    w1={mark=o,draw=col1,ultra thick,scale=1.6},
    w32={mark=o,draw=col3,ultra thick,scale=1.6},
    w64={mark=o,draw=col5,ultra thick,scale=1.6},
    w128={mark=o,draw=col6,ultra thick,scale=1.6},
    w256={mark=o,draw=col7,ultra thick,scale=1.6},
    c1={mark=x,draw=col1,ultra thick,scale=1.5},
    c32={mark=x,draw=col3,ultra thick,scale=1.5},
    c64={mark=x,draw=col5,ultra thick,scale=1.5},
    c128={mark=x,draw=col6,ultra thick,scale=1.5},
    c256={mark=x,draw=col7,ultra thick,scale=1.5}
    },
    hide axis,
    only marks,
    ymin=0, ymax=1,
    xmin=0, xmax=1,
]
\addplot+ [mark=*] coordinates { };
\addlegendimage{only marks,mark=square*,ultra thick,scale=1.6,col1}
\addlegendentry{1 Team}
\addlegendimage{only marks,mark=square*,ultra thick,scale=1.6,col3}
\addlegendentry{32 Teams}
\addlegendimage{only marks,mark=square*,ultra thick,scale=1.6,col5}
\addlegendentry{64 Teams}
\addlegendimage{only marks,mark=square*,ultra thick,scale=1.6,col6}
\addlegendentry{128 Teams}
\addlegendimage{only marks,mark=square*,ultra thick,scale=1.6,col7}
\addlegendentry{256 Teams}
\end{axis}

\end{tikzpicture} 
    }
    \vspace{-3mm}
    \caption{
    Comparison of the performance between the NVIDIA-provided \lstinline{malloc} and our domain-specific balanced allocator with 32 thread slots and 16 team slots (refer to ~\Cref{sec:allocators}).
    The benchmark is an exaggeration of the allocation scheme in some SPEC OMP benchmarks, where memory is allocated and deallocated at the beginning and end of a parallel region with all threads.
    }
    \Description{
    Comparison of the performance between the NVIDIA-provided \lstinline{malloc} and our domain-specific balanced allocator with 32 thread slots and 16 team slots (refer to ~\Cref{sec:allocators}).
    The benchmark is an exaggeration of the allocation scheme in some SPEC OMP benchmarks, where memory is allocated and deallocated at the beginning and end of a parallel region with all threads.
    }
    \label{fig:malloc}
\vspace{-3mm}
\end{figure}
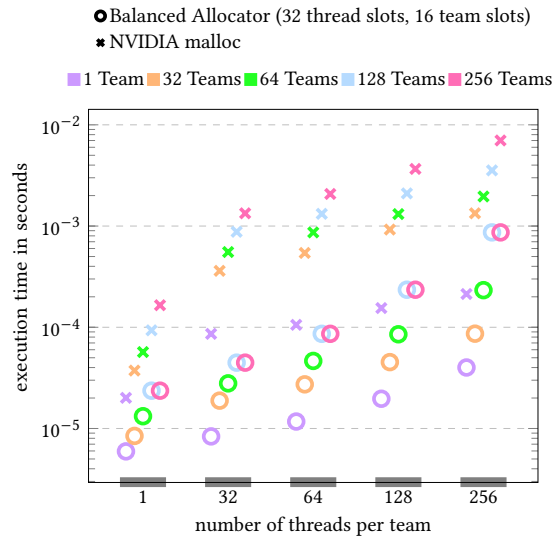

\subsection{RPC Performance}

To measure the overhead of an RPC call, we conducted a profiling experiment where we called the \lstinline{fprintf} function 1000 times with the following arguments: \lstinline{fprintf(stderr, "fread reads: 
In this example, \lstinline|buffer| points to a 128 byte array that has to be copied back and forth as the read/write behavior of \lstinline|fprintf| arguments is unknown (without inspecting the format string).
The average device time spend per RPC was 975 microseconds.
The distribution of this time spend is visualized in \Cref{fig:rpc}.
From left to right, the following stages are traversed on the device (top part):
\begin{inparaenum}[1)]
\item $0.1$\% of the overall time is spent on initializing the RPC argument information (\lstinline{RPCArgInfo} in \Cref{fig:rpc_gen_device}).
\item $9.1\%$ of the time is on identifying the underlying objects of the three pointer arguments, which includes copying the format string and \lstinline{buffer} to an RPC buffer where the host can access them (managed memory, as described in \Cref{sec:background}). 
\item The device thread spends $89$\% of the time waiting for the host to act on the request and acknowledge that it has been performed.
\item $1.8\%$ of the time is spent copying the data from the RPC buffer back to the \lstinline{buffer}.
\end{inparaenum}
On the host, the time is spent like this, again from left to right:
\begin{inparaenum}[1)]
\item $2$\% of the overall time is spent on copying the RPC information (\lstinline{RPCInfo} in \Cref{fig:rpc_gen_host}) to the host.
\item $3.5$\% of the time is spent invoking the host wrapper, which in turn calls the actual \lstinline{fprintf} function and sets up the return value.
\item $5.4$ of the time is spent on copying \lstinline{RPCInfo} object back to the device and notifying completion.
This notification is done by setting an integer value to 0 that is in managed memory and is also accessible to the device.
\item The remaining $89.1$\% of the time is the gap between when the host notifies completion and when the device receives the notification.
This gap occurs because threads in a GPU kernel that is already running are not guaranteed to see the updates to memory done by the CPU or other devices in a specific order and within a specific time interval \cite{DBLP:conf/hipc/PotluriGRNVI17,DBLP:conf/hpdc/DaoudWS16}.
\end{inparaenum}

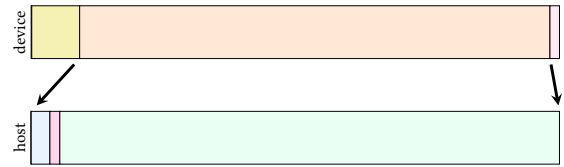
\begin{figure}[tb]
\lstset{basicstyle=\footnotesize\lst@ifdisplaystyle\small\linespread{0.94}\fi\ttfamily,
        }
    \centering
    \resizebox{.9\linewidth}{!}{
      \newcommand{\timebar}[8]{
\draw[fill=#5] ( 0, 0) node (#4-1bottom) {} rectangle (#1, 1) node (#4-1top) {};
\draw[fill=#6] (#1, 0) node (#4-2bottom) {} rectangle (#2, 1) node (#4-2top) {};
\draw[fill=#7] (#2, 0) node (#4-3bottom) {} rectangle (#3, 1) node (#4-3top) {};
\ifnum10>0#3\relax
\draw[fill=#8] (#3, 0) node (#4-4bottom) {}  rectangle (10, 1) node (#4-4top) {};
\fi
}

\begin{tikzpicture}

\timebar{0.01118569379}{0.917913822}{9.819808828}{dev}{col1!30}{col2!30}{col3!30}{col4!30}
\node[rotate=90, anchor=south] at (0,0.5) {device};

\tikzset{shift={(0,-2)}}

\timebar{0.000431613533}{0.355022151501}{0.542755359371}{host}{col5!30}{col6!30}{col7!30}{col8!30}
\node[rotate=90, anchor=south] at (0,0.5) {host};

\draw[ultra thick, ->] (dev-3bottom) -- (host-1top);
\draw[ultra thick, ->] (dev-4bottom) -- (host-4top);

\end{tikzpicture} 
    }
    \vspace{-3mm}
    \caption{Visualization of the time (avg. total 975 microseconds) spend in different staging while resolving a \lstinline|fprintf| RPC.}
    \Description{Visualization of the time (avg. total 975 microseconds) spend in different staging while resolving a \lstinline|fprintf| RPC.}
    \label{fig:rpc}
\vspace{-4mm}
\end{figure}

\subsection{Parallel Region Modeling}

In the following we compare the performance obtained via \emph{GPU First} compilation of CPU code against manual offload versions of various parallel regions to determine if the proposed methodology is suitable to guide porting efforts.

\begin{figure*}[htb]
\input{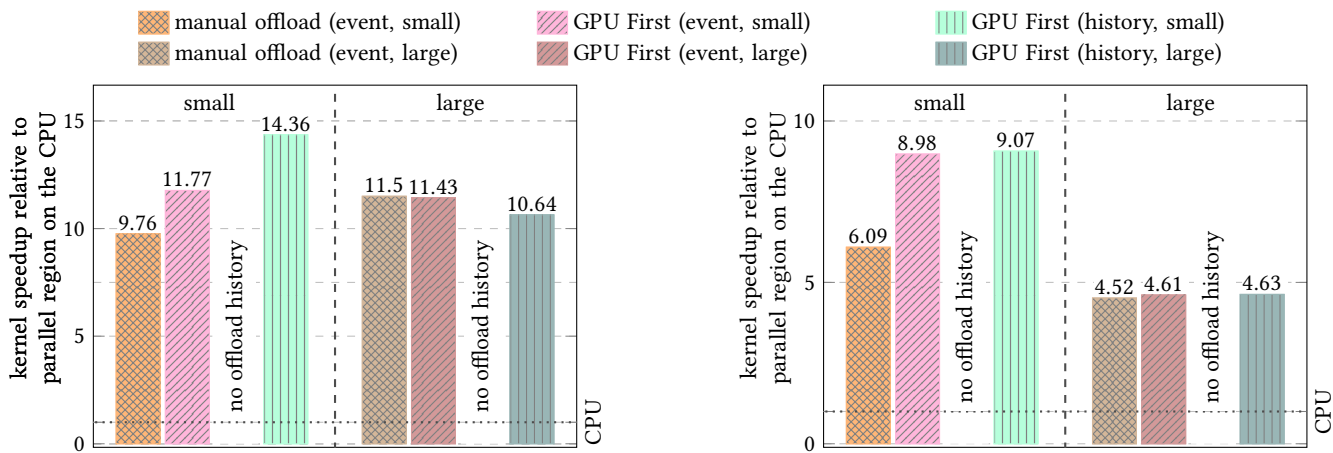}
\begin{minipage}[t]{.45\linewidth}\centering%
    \begin{tikzpicture}

\begin{axis}[
    enlarge y limits={abs=0.01},
    enlarge x limits={abs=0.5},
    xticklabels={},
    yticklabels={},
    legend columns=3,
    legend style={
      at={(1.33,1.13)},
      anchor=center,
      draw=none,
      /tikz/every even column/.append style   = {column sep= 0mm, text width=15em},
    },
    legend image code/.code={
      \draw [fill=#1] (0cm,-1.5mm) rectangle (4mm,1.5mm);
    },
    xmin=0, xmax=1,
    ymin=0.0, ymax=1.1,
    width=\linewidth,
    height=64mm,
    xmajorticks=false,
    axis line style={draw=none},
    tick style={draw=none},
]
\addlegendimage{col3, postaction={pattern color=gray, pattern=crosshatch}}
\addlegendentry{manual offload (event, small)}
\addlegendimage{col4, postaction={pattern color=gray, pattern=north east lines}}
\addlegendentry{GPU First (event, small)}
\addlegendimage{col8, postaction={pattern color=gray, pattern=vertical lines}}
\addlegendentry{GPU First (history, small)}
\addlegendimage{col10!40, postaction={pattern color=gray, pattern=crosshatch}}
\addlegendentry{manual offload (event, large)}
\addlegendimage{col11!40, postaction={pattern color=gray, pattern=north east lines}}
\addlegendentry{GPU First (event, large)}
\addlegendimage{col13!40, postaction={pattern color=gray, pattern=vertical lines}}
\addlegendentry{GPU First (history, large)}
\end{axis}

\begin{axis}[
    extra y ticks={1},
    extra y tick labels={CPU},
    ymin=0.0, ymax=16.5,
    name nodes near coords/.default=coordnode,
    bar width=0.10,
    name nodes near coords/.style={
      every node near coord/.append style={
        shift={(axis direction cs:0, -\ycoord)},
        name=#1-\coordindex,
        alias=#1-last,
        rotate=0,
        anchor=south,
        align=center,
        /pgf/number format/fixed,
      },
    },
]

\addplot+[
  name nodes near coords=offloadEventSmall,
  col3, postaction={pattern color=gray, pattern=crosshatch},
 ] 
coordinates {
  (0, 9.757)
  };

\addplot+[
  name nodes near coords=gpuFirstEventSmall,
  col4, postaction={pattern color=gray, pattern=north east lines},
]
coordinates {
  (0, 11.772)
  };

\addplot+[
  name nodes near coords=missing1,
  white,
]
coordinates {
(0,0)
  };

\addplot+[
  name nodes near coords=gpuFirstHistorySmall,
  col8, postaction={pattern color=gray, pattern=vertical lines},
]
coordinates {
  (0, 14.360)
  };

\addplot+[
]
coordinates {
  };

\addplot+[
  name nodes near coords=offloadEventLarge,
  col10!40, postaction={pattern color=gray, pattern=crosshatch},
]
coordinates {
  (0, 11.503)
  };

\addplot+[
  name nodes near coords=gpuFirstEventLarge,
  col11!40, postaction={pattern color=gray, pattern=north east lines},
]
coordinates {
  (0, 11.434)
  };

\addplot+[
  name nodes near coords=missing2,
  white,
]
coordinates {
(0,0)
  };

\addplot+[
  name nodes near coords=gpuFirstHistory,
  col13!40, postaction={pattern color=gray, pattern=vertical lines},
]
coordinates {
  (0, 10.64)
  };

\addplot[black!70,dotted,thick,sharp plot,nodes near coords={},
update limits=false,shorten >=-3mm,shorten <=-3mm]
coordinates {(-1,1) (1,1)};
\addplot[black!70,dashed,thick,sharp plot,nodes near coords={},
update limits=false,shorten >=-3mm,shorten <=-3mm]
coordinates {(0,0) (0,18)};
\node[] (top) at (axis description cs:1,1) {};
\node[anchor=north] (small) at (axis description cs:0.24,1) {small};
\node[anchor=north] (large) at (axis description cs:0.76,1) {large};

\end{axis}

\node[rotate=90,anchor=west,fill=white] (test) at ($(missing1-0) + (0,0.2)$) {no offload history};
\node[rotate=90,anchor=west,fill=white] (test) at ($(missing2-0) + (0,0.2)$) {no offload history};

\end{tikzpicture}
    \subcaption{
      Performance of the compute kernel of XSBench relative to the CPU version.
    }
    \label{fig:xbench}%
\end{minipage}%
\hfill%
\begin{minipage}[t]{.45\linewidth}\centering%
    \begin{tikzpicture}
\begin{axis}[
    extra y ticks={1},
    extra y tick labels={CPU},
    ymin=0.0, ymax=11,
    name nodes near coords/.default=coordnode,
    bar width=0.10,
    name nodes near coords/.style={
      every node near coord/.append style={
        shift={(axis direction cs:0, -\ycoord)},
        name=#1-\coordindex,
        alias=#1-last,
        rotate=0,
        anchor=south,
        align=center,
        /pgf/number format/fixed,
      },
    },
]

\addplot+[
  name nodes near coords=offloadEventSmall,
  col3, postaction={pattern color=gray, pattern=crosshatch},
 ] 
coordinates {
  (0, 6.090)
  };

\addplot+[
  name nodes near coords=gpuFirstEventSmall,
  col4, postaction={pattern color=gray, pattern=north east lines},
]
coordinates {
  (0, 8.976)
  };

\addplot+[
  name nodes near coords=missing1,
  white,
]
coordinates {
(0,0)
  };

\addplot+[
  name nodes near coords=gpuFirstHistorySmall,
  col8, postaction={pattern color=gray, pattern=vertical lines},
]
coordinates {
  (0, 9.0669)
  };

\addplot+[
]
coordinates {
  };

\addplot+[
  name nodes near coords=offloadEventLarge,
  col10!40, postaction={pattern color=gray, pattern=crosshatch},
]
coordinates {
  (0, 4.5159)
  };

\addplot+[
  name nodes near coords=gpuFirstEventLarge,
  col11!40, postaction={pattern color=gray, pattern=north east lines},
]
coordinates {
  (0, 4.6102)
  };

\addplot+[
  name nodes near coords=missing2,
  white,
]
coordinates {
(0,0)
  };

\addplot+[
  name nodes near coords=gpuFirstHistory,
  col13!40, postaction={pattern color=gray, pattern=vertical lines},
]
coordinates {
  (0, 4.6277)
  };

\addplot[black!70,dotted,thick,sharp plot,nodes near coords={},
update limits=false,shorten >=-3mm,shorten <=-3mm]
coordinates {(-1,1) (1,1)};
\addplot[black!70,dashed,thick,sharp plot,nodes near coords={},
update limits=false,shorten >=-3mm,shorten <=-3mm]
coordinates {(0,0) (0,18)};
\node[] (top) at (axis description cs:1,1) {};
\node[anchor=north] (small) at (axis description cs:0.24,1) {small};
\node[anchor=north] (large) at (axis description cs:0.76,1) {large};

\end{axis}

\node[rotate=90,anchor=west,fill=white] (test) at ($(missing1-0) + (0,0.32)$) {no offload history};
\node[rotate=90,anchor=west,fill=white] (test) at ($(missing2-0) + (0,0.32)$) {no offload history};

\end{tikzpicture}
    \subcaption{
      Performance of the compute kernel of RSBench relative to the CPU version.
    }
    \label{fig:rsbench}%
\end{minipage}
    \vspace*{-3mm}

\caption{Performance of different GPU versions of the OpenMC proxy applications XSBench and RSBench compared to their respective CPU counterpart.}
\Description{Performance of different GPU versions of the OpenMC proxy applications XSBench and RSBench compared to their respective CPU counterpart.}
\vspace*{-3mm}
\label{fig:xsrsbench}
    
\end{figure*}


\subsubsection{XSBench and RSBench}

The two OpenMC proxy applications XSBench~\cite{XS_Tramm_2014} (v20) and RSBench~\cite{DBLP:conf/easc/TrammSFJ14} (v13) are implemented in different parallel programming models, including OpenMP threading for the CPU, and OpenMP offload for the GPU.
In the former, two alternative methods are available to perform the cross-section lookup as part of the neutron transport simulation: event-based lookup and history-based lookup.
In the offloading version history-based mode was not implemented but we can test it out with the \emph{GPU First} methodology using the CPU implementation.
The results for both benchmarks and two different input sizes are shown in \Cref{fig:xbench} and \Cref{fig:rsbench}, respectively.
For the small input size, history mode is actually outperforming the event mode on the GPU.
However, with the large input size event mode has caught up (RSBench), or even surpassed (XSBench), history mode.
These results validate the choice of event-based mode for the offloading implementation.
The second insight from the evaluation can be derived by comparing the event-based results obtained via the manually offloaded version and the \emph{GPU First} version.
For the small input the \emph{GPU First} versions are likely to benefit from cache re-use as the data initialization is also performed on the GPU.
However, with the large input the two results are a close match.
Thus, performance predictions obtained via \emph{GPU First} and the original CPU-only version would have provided accurate guidance for a potential manual port to the GPU.

\subsubsection{HeCBench: Interleaved}

The HeCBench~\cite{zjin-lcf/HeCBench} ``interleaved'' micro benchmark originated from \citet{cook2012cuda} and shows how different memory access patterns behave on the CPU and GPU.
We timed the parallel region with interleaved memory accesses (array-of-struct inputs) as well as the one with non-interleaved accesses (struct-of-array inputs) on the CPU and GPU.
The results, expressed as speedups and slowdowns of the GPU version, are shown in \Cref{fig:interleaved}.
While the \emph{GPU First} version shows the same tendency as the manually offloading version, we needed to explicitly match the number of teams to perfectly match the result with our automatically offloaded parallel regions.

\begin{figure}[htbp]
\input{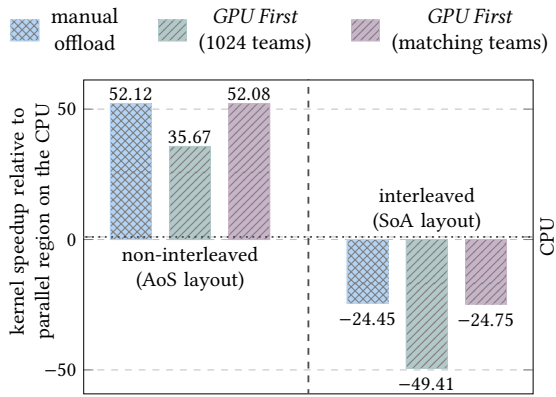}
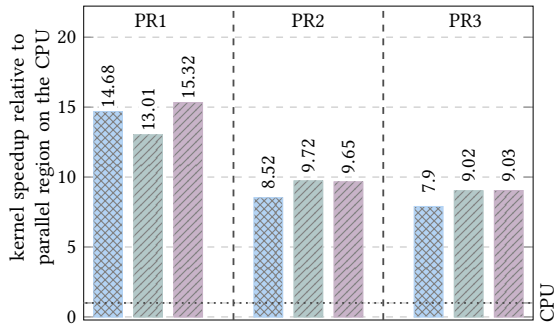
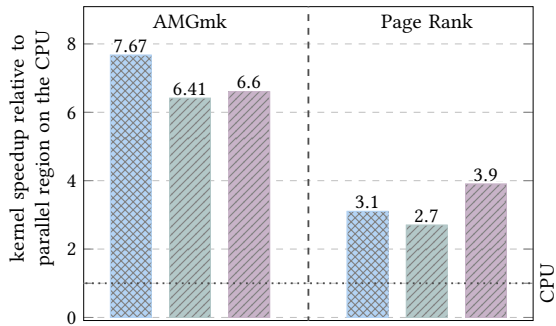

\begin{minipage}[t]{\linewidth}\centering%
    \resizebox{.9\linewidth}{!}{
    \begin{tikzpicture}
\begin{axis}[
    legend columns=3,
    legend style={
      at={(0.5,1.15)},
      anchor=center,
      draw=none,
      /tikz/every even column/.append style   = {column sep= 5mm},
      /tikz/every odd column/.append style   = {column sep= 1mm},
    },
    legend image code/.code={
      \draw [fill=#1] (0cm,-1.5mm) rectangle (4mm,1.5mm);
    }, 
    legend style={cells={align=center}},
    xmin=0, xmax=1,
    ymin=0.0, ymax=1.1,
    width=\linewidth,
    height=64mm,
    hide axis,    
]
\addlegendimage{
  col12!30, postaction={pattern color=gray, pattern=crosshatch}}
\addlegendentry{manual\\ offload}
\addlegendimage{
  col13!30, postaction={pattern color=gray, pattern=north east lines}}
\addlegendentry{\emph{GPU First}\\ (1024 teams)}
\addlegendimage{
  col14!30, postaction={pattern color=gray, pattern=north east lines}}
\addlegendentry{\emph{GPU First}\\ (matching teams)}
\end{axis}
\end{tikzpicture}
}
\end{minipage}
\vspace{-2mm}

\begin{minipage}[t]{\linewidth}\centering%
    \resizebox{.9\linewidth}{!}{
      \begin{tikzpicture}
\begin{axis}[
    extra y ticks={1},
    extra y tick labels={CPU},
    ymin=-58, ymax=60,
    name nodes near coords/.default=coordnode,
    bar width=0.10,
    name nodes near coords/.style={
      every node near coord/.append style={
        shift={(axis direction cs:0, -\ycoord)},
        name=#1-\coordindex,
        alias=#1-last,
        rotate=0,
        anchor=south,
        align=center,
        /pgf/number format/fixed,
      },
    },
    ybar=8pt,
]

\addplot+[
  name nodes near coords=niOffload,
  col12!30, postaction={pattern color=gray, pattern=crosshatch},
 ] 
coordinates {
  (0, 52.12090036)
  };

\addplot+[
  name nodes near coords=niMT1024,
  col13!30, postaction={pattern color=gray, pattern=north east lines},
 ] 
coordinates {
  (0, 35.67445016)
  };

\addplot+[
  name nodes near coords=niMT16,
  col14!30, postaction={pattern color=gray, pattern=north east lines},
 ] 
coordinates {
  (0, 52.08492271)
  };

\addplot+[
 ] 
coordinates {
  };

\addplot+[
  name nodes near coords=iOffload,
  col12!30, postaction={pattern color=gray, pattern=crosshatch},
]
coordinates {
  (0, -24.45384987)
  };
  
\addplot+[
  name nodes near coords=iMT1024,
  col13!30, postaction={pattern color=gray, pattern=north east lines},
 ] 
coordinates {
  (0, -49.40655542)
  };

\addplot+[
  name nodes near coords=iMT16,
  col14!30, postaction={pattern color=gray, pattern=north east lines},
 ] 
coordinates {
  (0, -24.74615046)
  };

\addplot[black!70,dotted,thick,sharp plot,nodes near coords={},
update limits=false,shorten >=-3mm,shorten <=-3mm]
coordinates {(-1,1) (1,1)};
\addplot[black!70,dashed,thick,sharp plot,nodes near coords={},
update limits=false,shorten >=-3mm,shorten <=-3mm]
coordinates {(0,-200) (0, 200)};
\node[] (top) at (axis description cs:1,1) {};
\node[] (middle) at (0,0) {};

\end{axis}

\ExtractCoordinateA{middle}
\ExtractCoordinateB{niMT1024-0}
\node[anchor=north, text width=25mm,align=center] (noninterleaved) at (\XCoordB, \YCoordA) {non-interleaved (AoS layout)};
\ExtractCoordinateB{iMT1024-0}
\node[anchor=south, text width=25mm,align=center] (interleaved) at (\XCoordB, \YCoordA) {interleaved (SoA layout)};

\end{tikzpicture} 
    }
    \subcaption{Relative performance of the two parallel regions in the interleaved benchmark (from HeCBench) when executed on the GPU instead of the CPU. The first (non-interleaved) uses a struct-of-array layout while the second (interleaved) uses an array-of-struct layout.}
    \label{fig:interleaved}
\end{minipage}
\vspace{3mm}

\begin{minipage}[t]{\linewidth}\centering%
    \resizebox{.9\linewidth}{!}{
      \begin{tikzpicture}
\begin{axis}[
    extra y ticks={1},
    extra y tick labels={CPU},
    ymin=0, ymax=22,
    name nodes near coords/.default=coordnode,
    bar width=0.07,
    name nodes near coords/.style={
      every node near coord/.append style={
        shift={(axis direction cs:0, -\ycoord)},
        name=#1-\coordindex,
        alias=#1-last,
        rotate=0,
        anchor=south,
        align=center,
        /pgf/number format/fixed,
      },
    },
    ybar=5pt,
      nodes near coords = {%
      \begingroup%
        \pgfkeys{/pgf/fpu}%
        \pgfmathparse{\pgfplotspointmeta<0.0}%
        \global\let\isneg=\pgfmathresult%
        \pgfmathparse{\pgfplotspointmeta==0}%
        \global\let\iszero=\pgfmathresult%
        \pgfmathparse{\pgfplotspointmeta==1}%
        \global\let\isone=\pgfmathresult%
      \endgroup%
      \pgfmathfloatcreate{0}{0.0}{0}%
      \let\ZERO=\pgfmathresult%
      \pgfmathfloatcreate{1}{1.0}{0}%
      \let\ONE=\pgfmathresult%
      \pgfmathfloatcreate{1}{-1.0}{0}%
      \let\MONE=\pgfmathresult%
      \color{black}%
      \global\let\ycoord=\ZERO
    \ifx\iszero\ONE%
    \else%
        \ifx\isneg\ONE%
        {%
          \node[rotate=90,fill=white,yshift=-4pt,xshift=9pt,inner sep=1pt]{\color{black}\textbf{\pgfmathprintnumber[precision=2]{\rawy}}};%
        }%
        \else%
          \node[rotate=90,fill=white,yshift=-4pt,xshift=9pt,inner sep=1pt]{\color{black}\textbf{\pgfmathprintnumber[precision=2]{\rawy}}};%
        \fi%
    \fi%
    },
]

\addplot+[
  name nodes near coords=off1,
  col12!30, postaction={pattern color=gray, pattern=crosshatch},
 ] 
coordinates {
  (0, 14.67804854)
  };

\addplot+[
  name nodes near coords=mt1,
  col13!30, postaction={pattern color=gray, pattern=north east lines},
 ] 
coordinates {
  (0, 13.0136894)
  };
\addplot+[
  name nodes near coords=mt1match,
  col14!30, postaction={pattern color=gray, pattern=north east lines},
 ] 
coordinates {
  (0, 15.32424341)
  };

\addplot+[
 ] 
coordinates {
  };

\addplot+[
  name nodes near coords=off2,
  col12!30, postaction={pattern color=gray, pattern=crosshatch},
 ] 
coordinates {
  (0, 8.522244067)
  };
\addplot+[
  name nodes near coords=mt2,
  col13!30, postaction={pattern color=gray, pattern=north east lines},
]
coordinates {
  (0, 9.723230855)
  };
\addplot+[
  name nodes near coords=mt2match,
  col14!30, postaction={pattern color=gray, pattern=north east lines},
 ] 
coordinates {
  (0, 9.65147391)
  };

\addplot+[
 ] 
coordinates {
  };

\addplot+[
  name nodes near coords=off3,
  col12!30, postaction={pattern color=gray, pattern=crosshatch},
 ] 
coordinates {
  (0, 7.895095135)
  };

\addplot+[
  name nodes near coords=mt3,
  col13!30, postaction={pattern color=gray, pattern=north east lines},
 ] 
coordinates {
(0, 9.015701821)
  };
\addplot+[
  name nodes near coords=mt3match,
  col14!30, postaction={pattern color=gray, pattern=north east lines},
 ] 
coordinates {
  (0, 9.034509518)
  };

\addplot[black!70,dotted,thick,sharp plot,nodes near coords={},
update limits=false,shorten >=-3mm,shorten <=-3mm]
coordinates {(-1,1) (1,1)};

\draw[black!70,dashed,thick,] (axis description cs:0.3333,0) -- (axis description cs:0.3333, 1);
\draw[black!70,dashed,thick,] (axis description cs:0.6666,0) -- (axis description cs:0.6666, 1);
\node[] (top) at (axis description cs:1,1) {};
\node[] (middle) at (0,0) {};

\node[anchor=north] (pr1) at (axis description cs:0.15, 1) {PR1};
\node[anchor=north] (pr2) at (axis description cs:0.5, 1) {PR2};
\node[anchor=north] (pr3) at (axis description cs:0.85, 1) {PR3};

\end{axis}

\end{tikzpicture} 
    }
    \subcaption{Relative performance of the three parallel regions (PR1, PR2, PR3) in the hypterm micro benchmark (from HeCBench) when executed on the GPU instead of the CPU.}
    \label{fig:hypeterm}
\end{minipage}
\vspace{3mm}

\begin{minipage}[t]{\linewidth}\centering%
    \resizebox{.9\linewidth}{!}{
      \begin{tikzpicture}
\begin{axis}[
    extra y ticks={1},
    extra y tick labels={CPU},
    ymin=0.0, ymax=9,
    name nodes near coords/.default=coordnode,
    bar width=0.10,
    name nodes near coords/.style={
      every node near coord/.append style={
        shift={(axis direction cs:0, -\ycoord)},
        name=#1-\coordindex,
        alias=#1-last,
        rotate=0,
        anchor=south,
        align=center,
        /pgf/number format/fixed,
      },
    },
    ybar=8pt,
]

\addplot+[
  name nodes near coords=amgmkOffload,
  col12!30, postaction={pattern color=gray, pattern=crosshatch},
 ] 
coordinates {
  (0, 7.670604372)
  };

\addplot+[
  name nodes near coords=amgmkMT,
  col13!30, postaction={pattern color=gray, pattern=north east lines},
]
coordinates {
  (0, 6.410588883)
  };
  
\addplot+[
  name nodes near coords=amgmkMTmatch,
  col14!30, postaction={pattern color=gray, pattern=north east lines},
]
coordinates {
  (0, 6.601275915)
  };
  
\addplot+[
 ] 
coordinates {
  };

\addplot+[
  name nodes near coords=pageRankOffload,
  col12!30, postaction={pattern color=gray, pattern=crosshatch},
 ] 
coordinates {
  (0, 3.102152887)
  };

\addplot+[
  name nodes near coords=pageRankMT,
  col13!30, postaction={pattern color=gray, pattern=north east lines},
]
coordinates {
  (0, 2.698591605)
  };

\addplot+[
  name nodes near coords=pageRankMTMatch,
  col14!30, postaction={pattern color=gray, pattern=north east lines},
 ] 
coordinates {
  (0, 3.900568937)
  };

\addplot[black!70,dotted,thick,sharp plot,nodes near coords={},
update limits=false,shorten >=-3mm,shorten <=-3mm]
coordinates {(-1,1) (1,1)};
\addplot[black!70,dashed,thick,sharp plot,nodes near coords={},
update limits=false,shorten >=-3mm,shorten <=-3mm]
coordinates {(0,0) (0,18)};
\node[] (top) at (axis description cs:1,1) {};

\end{axis}

\ExtractCoordinateA{top}
\ExtractCoordinateB{amgmkMT-0}
\node[anchor=north, text width=25mm,align=center] (amg) at (\XCoordB, \YCoordA) {AMGmk};
\ExtractCoordinateB{pageRankMT-0}
\node[anchor=north, text width=25mm,align=center] (pr) at (\XCoordB, \YCoordA) {Page Rank};

\end{tikzpicture} 
    }
    \subcaption{Relative performance results for the timed parallel regions in the AMGmk and page-rank micro benchmark (from HeCBench) when executed on the GPU instead of the CPU.}
    \label{fig:mixed}
\end{minipage}

\vspace{-2mm}
\caption{
Comparison of micro benchmarks performance results for a parallel region compiled with \emph{GPU First} to the GPU, and the manually offloaded counterpart, relative to the corresponding CPU parallel region.
The matching teams column for \emph{GPU First} uses the same number of teams as the manually offloaded version.
The legend at the top of the figure is shared among all plots.
}
\Description{
Comparison of micro benchmarks performance results for a parallel region compiled with \emph{GPU First} to the GPU, and the manually offloaded counterpart, relative to the corresponding CPU parallel region.
The matching teams column for \emph{GPU First} uses the same number of teams as the manually offloaded version.
The legend at the top of the figure is shared among all plots.
}
\label{fig:AAA}
    
\end{figure}

\subsubsection{HeCBench: Hypterm}

The HeCBench ``hypterm'' micro benchmark is a complex stencil operation that originated from the ExpCNS Compressible Navier-Stokes mini-application~\cite{ExaCT} and was extracted by \citet{DBLP:conf/ppopp/RawatRSPRS18}.
The GPU version in in HeCBench contains three kernels which we transformed into three parallel regions for the CPU.
The results of the GPU version and the \emph{GPU First} version relative to the CPU are shown in \Cref{fig:hypeterm}.
While the original GPU version is slightly slower, the overall performance behavior matches the \emph{GPU First} prediction.

\subsubsection{HeCBench: AMGmk, Page-Rank, }

\Cref{fig:mixed} shows the results obtained for the AMGmk, page-rank benchmarks.
The first measures only the relax kernel of the original AMGmk proxy application~\cite{CORAL}.
The second is an implementation of the page-rank algorithm for graphs in which the propagation step is measured.

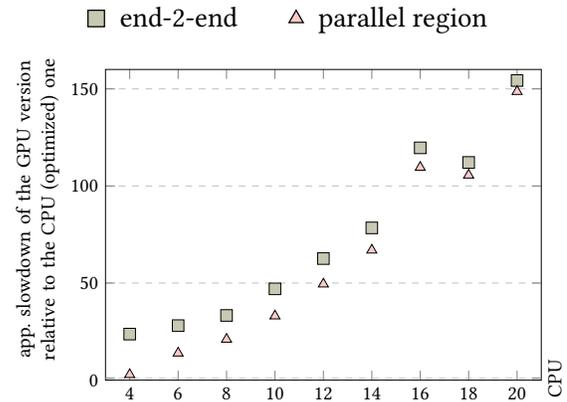
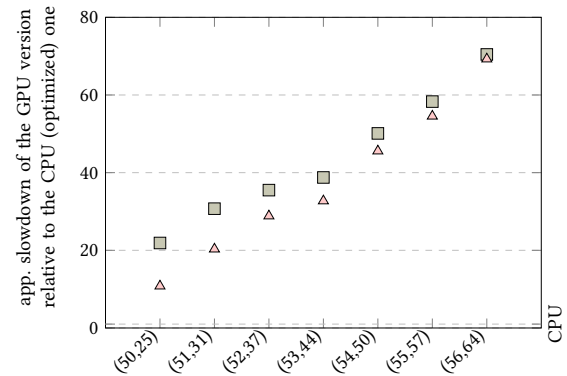
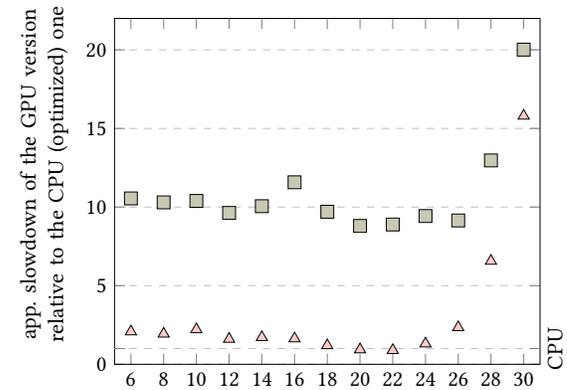
\begin{figure}

\begin{minipage}[t]{\linewidth}\centering%
    \resizebox{.7\linewidth}{!}{
    \begin{tikzpicture}
\begin{axis}[
    legend columns=2,
    legend style={
      at={(0.5,1.15)},
      anchor=center,
      draw=none,
      /tikz/every even column/.append style   = {column sep= 5mm},
      /tikz/every odd column/.append style   = {column sep= 1mm},
    },
    legend style={cells={align=center}},
    xmin=0, xmax=1,
    ymin=0.0, ymax=1.1,
    width=\linewidth,
    height=64mm,
    hide axis,    
]
\addlegendimage{  only marks,
  every mark/.append style={solid, draw=black, fill=col15!30,scale=1.3}, 
  mark=square*}
\addlegendentry{end-2-end}
\addlegendimage{  only marks,
  every mark/.append style={solid, draw=black, fill=col16!30,scale=1.3}, 
  mark=triangle*}
\addlegendentry{parallel region}
\end{axis}
\end{tikzpicture}
}
\end{minipage}
\vspace{-4mm}

\begin{minipage}[t]{\linewidth}\centering%
    \resizebox{.9\linewidth}{!}{
      \begin{tikzpicture}

\begin{axis}[
    extra y ticks={1},
    extra y tick labels={CPU},
    ymin=0.0, ymax=160,
    name nodes near coords/.default=coordnode,
    bar width=0.10,
    name nodes near coords/.style={
      every node near coord/.append style={
        shift={(axis direction cs:0, -\ycoord)},
        name=#1-\coordindex,
        alias=#1-last,
        rotate=0,
        anchor=south,
        align=center,
        /pgf/number format/fixed,
      },
    },
    xmin=4,  xmax=20,
    xtick distance={2},
    enlarge x limits={abs=1},
    width=\linewidth,
    height=65mm,
     extra y tick style={
           grid=major,
           tick label style={rotate=90},
           ticklabel pos=right,
     },
    ylabel={app. slowdown of the GPU version relative to the CPU (optimized) one},
    ylabel style = {align=center,at={(-0.09,0.55)},text width=5cm},
]

\addplot+[
  name nodes near coords=end2end,
  only marks,
  every mark/.append style={solid, draw=black, fill=col15!30,scale=1.3}, 
  mark=square*,
 ] 
coordinates {
(4, 23.71291305)
(6, 28.08102772)
(8, 33.31944034)
(10, 47.00821984)
(12, 62.60580386)
(14, 78.4042739)
(16, 119.6299607)
(18, 112.1189084)
(20, 154.3024754)
  };

\addplot+[
  name nodes near coords=parRegion,
  only marks,
  every mark/.append style={solid, draw=black, fill=col16!30,scale=1.3}, 
  mark=triangle*,
 ] 
coordinates {
(4,2.853754473) 
(6, 13.90612513)
(8,21.02651895)
(10, 33.02571654)
(12, 49.47597231) 
(14, 66.95620005)
(16, 109.5703832)
(18, 105.5760577)
(20, 148.6172162)
  };

\end{axis}
\end{tikzpicture} 
    }
\vspace{-2mm}
    \subcaption{358.botsalgn. The x axis is the number of input sequences.
    This benchmark distributes sequences across multiple threads through an outer \lstinline{parallel} region, where each thread spawns several OpenMP tasks to execute the pair alignment algorithm.}
    \label{fig:spec-358}
\end{minipage}
\begin{minipage}[t]{\linewidth}\centering%
    \resizebox{.9\linewidth}{!}{
      \begin{tikzpicture}

\begin{axis}[
    extra y ticks={1},
    extra y tick labels={CPU},
    ymin=0.0, ymax=80,
    name nodes near coords/.default=coordnode,
    bar width=0.10,
    name nodes near coords/.style={
      every node near coord/.append style={
        shift={(axis direction cs:0, -\ycoord)},
        name=#1-\coordindex,
        alias=#1-last,
        rotate=0,
        anchor=south,
        align=center,
        /pgf/number format/fixed,
      },
    },
    xmin=1,  xmax=7,
    xtick distance={1},
    xtick={1,2,3,4,5,6,7},
    xticklabels={
(50,25), 
(51,31),
(52,37),
(53,44),
(54,50),
(55,57),
(56,64)
},
    enlarge x limits={abs=1},
    width=\linewidth,
    height=65mm,
     extra y tick style={
           grid=major,
           tick label style={rotate=90},
           ticklabel pos=right,
     },
    ylabel={app. slowdown of the GPU version relative to the CPU (optimized) one},
    ylabel style = {align=center,at={(-0.09,0.55)},text width=5cm},
    x tick label style={rotate=45,anchor=east},
]

\addplot+[
  name nodes near coords=end2end,
  only marks,
  every mark/.append style={solid, draw=black, fill=col15!30,scale=1.3}, 
  mark=square*,
 ] 
coordinates {
(1, 21.88757774) 
(2, 30.72475864)
(3, 35.5013504)
(4, 38.75906206)
(5, 50.10372313)
(6, 58.30036732)
(7, 70.4378462)
  };

\addplot+[
  name nodes near coords=parRegion,
  only marks,
  every mark/.append style={solid, draw=black, fill=col16!30,scale=1.3}, 
  mark=triangle*,
 ] 
coordinates {
(1, 10.79370488)
(2, 20.3155239)
(3, 28.8395581)
(4, 32.71760054)
(5, 45.56800443)
(6, 54.54240534)
(7, 69.26114314)
  };

\end{axis}
\end{tikzpicture} 
    }
\vspace{-3mm}
\subcaption{359.botsspar uses one thread creates tasks while the other threads in the parallel region execute them.
The x axis is the size of matrix and submatrix used in the benchmark.}
\label{fig:spec-359}
\end{minipage}
\begin{minipage}[t]{.9\linewidth}\centering%
    \resizebox{\linewidth}{!}{
      \begin{tikzpicture}

\begin{axis}[
    extra y ticks={1},
    extra y tick labels={CPU},
    ymin=0.0, ymax=22,
    name nodes near coords/.default=coordnode,
    bar width=0.10,
    name nodes near coords/.style={
      every node near coord/.append style={
        shift={(axis direction cs:0, -\ycoord)},
        name=#1-\coordindex,
        alias=#1-last,
        rotate=0,
        anchor=south,
        align=center,
        /pgf/number format/fixed,
      },
    },
    xmin=6,  xmax=30,
    xtick distance={2},
    enlarge x limits={abs=1},
    width=\linewidth,
    height=65mm,
     extra y tick style={
           grid=major,
           tick label style={rotate=90},
           ticklabel pos=right,
     },
    ylabel={app. slowdown of the GPU version relative to the CPU (optimized) one},
    ylabel style = {align=center,at={(-0.09,0.55)},text width=5cm},
]

\addplot+[
  name nodes near coords=offloadEventSmall,
  only marks,
  every mark/.append style={solid, draw=black, fill=col15!30,scale=1.3}, 
  mark=square*,
 ] 
coordinates {
(2,10.83923661) 
(4,6.43278232)
(6,10.54924858)
(8,10.29753388)
(10,10.39138599)
(12,9.626968046)
(14,10.05446202)
(16,11.57487804)
(18,9.701438092)
(20,8.800448367)
(22,8.890264957)
(24,9.428948964)
(26,9.147011498)
(28,12.96558124)
(30,20.01402426)
  };

\addplot+[
  name nodes near coords=parRegion,
  only marks,
  every mark/.append style={solid, draw=black, fill=col16!30,scale=1.3}, 
  mark=triangle*,
 ] 
coordinates {
(2 ,1.98299648 ) 
(4 ,0.037775978)
(6 ,2.076678019)
(8 ,1.934647426)
(10,2.212845429)
(12,1.606530646)
(14,1.714703073)
(16,1.633372769)
(18,1.198302583)
(20,0.935884944)
(22,0.891361062)
(24,1.308900685)
(26,2.346619088)
(28,6.567314424)
(30,15.79918583)
  };

\end{axis}
\end{tikzpicture} 
    }
\vspace{-5mm}
\subcaption{372.smithwa. The x axis is the sequence length.
The workload is manually distributed among multiple threads and threads communicate with each other using a producer-consumer model via shared variables followed by barriers.}
\label{fig:spec-372}
\end{minipage}

\vspace{-2mm}
\caption{
Relative performance results for the end-to-end execution and timed parallel regions in the three SPEC OMP 2012 benchmarks when executed on the GPU instead of the CPU.
The legend at the top is shared among all plots.}
\Description{
Relative performance results for the end-to-end execution and timed parallel regions in the three SPEC OMP 2012 benchmarks when executed on the GPU instead of the CPU.
The legend at the top is shared among all plots.}
\label{fig:spec}
\end{figure}

\subsubsection{SPEC OMP: 358.botsalgn and 359.botsspar}
These are two task-based benchmarks~\cite{DBLP:conf/icpp/DuranTFMA09} from the SPEC OMP 2012 suite~\cite{DBLP:conf/iwomp/MullerBBFHHJMPRSWWK12}.
The former performs sequence alignment while the latter is a sparse LU decomposition.
They are parallelized with different OpenMP tasking strategies.
\Cref{fig:spec-358,fig:spec-359} shows the performance of \emph{GPU First} relative to the CPU.
Since LLVM/OpenMP does not support tasking on GPUs, tasks are executed immediately by the encountering thread.
This limitation severely affects the GPU performance of these benchmarks.

In the case of \texttt{358.botsalgn}, sequences are distributed across multiple threads through an outer \lstinline{parallel} region.
Each thread spawns several tasks to that perform the alignment.
Since the number of sequences is smaller than the number of CPU cores, threads not involved in the work sharing can execute the spawned tasks concurrently.
However, on the GPU only a small number of threads (equal to the number of sequences) are executing concurrently.

Similarly, in \lstinline{359.botsspar}, one thread creates tasks while the other threads in the parallel region execute them.
This pattern of execution is equivalent to serial execution in our approach.
To enable parallelism for this benchmark, we rewrote the task regions by removing the \lstinline{task} construct and adding a \lstinline{parallel for} construct on the outer parallel region.
The results shown in \Cref{fig:spec-359} represent the threaded parallelism version of the benchmark.
The observed slowdown can be attributed to the lack of sufficient sequences to fully exploit the massive parallelism that GPUs offer, similar to the issue observed in benchmark \texttt{358.botsalgn}.
Nevertheless, our \emph{GPU First} scheme allows application developers to explore different parallelism on GPUs without much burden.

It is important to note that the lack of tasking support is not a limitation of our proposed scheme, but rather a limitation of the current LLVM/OpenMP implementation for GPUs.
If tasking is properly supported on the GPU, and there are a sufficient number of sequences, the massive parallelism of a GPU has the potential to make up for the performance difference between a CPU and a GPU thread.
While this means advancements in GPU tasking support could in the future improve performance of these codes on the GPU, the current results clearly indicate that a GPU port would require a different parallelization strategy.

\subsubsection{SPEC OMP: 372.smithwa}
\texttt{372.smithwa} implements the Smith-Waterman algorithm for sequence alignment and is characterized by a large number of nested loops with indirect memory accesses.
In this benchmark, the workload is first distributed to multiple threads, which maps well to the GPU.
However, the threads communicate with each other using a producer-consumer model via shared variables followed by barriers.
This form of communication is conceptually inefficient on GPUs.
\Cref{fig:spec-372} shows the performance of the \emph{GPU First} approach relative to the CPU.
As the input size is increased the relative performance is at first stable, indicating good scalability on the GPU.
However, when the sequence length hits 26 we can observe exponentially growing slowdown compared to the CPU execution.
Consequently, this benchmark is another example of an algorithm that is inefficient on the GPU, requiring a rewrite as part of the porting effort.
It is worth to note that without the balanced allocator the performance is dominated by the massively parallel allocations and deallocations at the beginning and end of the parallel region, respectively.

\section{Related Works}
\label{sec:related-works}

\subsection{GPU Execution of CPU Programs}
Several prior works have explored the execution of host programs on GPUs, including \citet{DBLP:conf/asplos/SilbersteinFKW13}, who proposed direct access to the host's file system from GPU code and implemented an RPC protocol to manage data transfers between the CPU and GPU.
\citet{DBLP:conf/date/DamschenRVP15} investigated transparent acceleration of binary applications using heterogeneous computing resources, without the need for manual porting or developer-provided hints.
Meanwhile, \citet{dblp:conf/cgo/MatsumuraZWEM20} proposed an automated stencil framework that can automatically transform and optimize stencil patterns in a given C source code, and generate corresponding CUDA code.
These works mainly focused on identifying and/or generating parts of the host program to run on GPUs.

\citet{DBLP:conf/ipps/MikushinLZB14} introduced a parallelization framework that detects parallelism and generates target code for both X86 CPUs and NVIDIA GPUs.
To support functions that can not be natively executed on GPUs, they replaced function calls in LLVM with an interface that ultimately results in the host executing the requested function using a foreign function interface (FFI).
However, our approach differs in two ways.
First, instead of relying on FFI, our compiler transformation generates the host wrapper, which restores the call site on the host.
Second, in their framework, GPU addresses are used directly on the host, which leads to segmentation faults when the host tries to access an address.
A signal handler for segmentation faults maps the GPU memory pages into CPU tables and copies input data.
However, this memory management subsystem does not work if the memory buffer is on the stack, such as when a local variable is used in host RPCs.
\citet{DBLP:conf/pldi/JablinPJJBA11} proposed a system for managing and optimizing CPU-GPU communication that is fully automatic.
Their system includes a run-time library and a set of compiler transformations that work together to manage and optimize communication between the CPU and GPU.
Unlike other approaches, this system does not rely on strength of static compile-time analyses or on programmer-supplied annotations.
Our pointer argument analysis shares a similar design to their work.

\citet{DBLP:conf/llvmhpc/TianHPCD22} were the first to attempt to run the entire host program on a GPU.
They proposed using OpenMP target offloading to leverage the portability of compiling and running host applications on a GPU.
However, their approach requires application developers to provide the wrapper function on both the host and device side, either manually or through scripts.
They also do not support variadic functions, and their paper shows severe performance regression due to single-team execution limitations.

\subsection{OpenMP Target Offloading}

In recent years, researchers have explored compiler and runtime optimization for OpenMP since OpenMP 4.0 introduced target offloading.
Bertolli et al. presented two works \cite{DBLP:conf/sc/BertolliAEOSJCS14,DBLP:conf/sc/BertolliABJECSS15} that enabled OpenMP offloading to GPUs in LLVM.
Flang, the PGI Fortran front-end, also supports OpenMP offloading via the LLVM OpenMP runtime~\cite{OzenAtzeniWolfeEtAlOMPGPUoffloadingFlang2018}. 
\citet{DBLP:conf/sc/AntaoBJBERMJOSC16} introduced front-end-based optimizations for Nvidia GPUs that can reduce register usage and avoid idle threads.
\citet{DBLP:conf/iwomp/DoerfertDF19} presented the TRegion interface, which supports more kernels to execute in SPMD mode.
\citet{DBLP:conf/lcpc/TianDC20} introduced runtime support for concurrent execution of OpenMP target tasks.
\citet{DBLP:conf/icppw/YviquelPFVLRCCD22} presented a framework for using the OpenMP programming model in distributed memory environments.
It provides a way to program clusters of shared memory machines with a hybrid approach that combines OpenMP directives and MPI communication.
\citet{DBLP:conf/cgo/HuberCGTDDCD22} presented OpenMP-aware program analyses and optimizations that allow efficient execution of CPU-centric parallelism on GPUs.
\citet{DBLP:conf/cc/OzenW22} introduced a fully descriptive model and demonstrate its benefits with an implementation of the \lstinline{loop} directive on NVIDIA GPUs.
\citet{DBLP:conf/ipps/DoerfertPHTDCG22} presented a co-design methodology for optimizing applications using an OpenMP GPU runtime with near-zero overhead, on top of which our device side host RPC support and partial \texttt{libc} implementation were built.

\section{Conclusion}
\label{sec:conclusion}

In this paper, we introduced a novel compilation scheme called ``\emph{GPU First}'' that enables automatic compilation of legacy CPU applications directly for GPUs without requiring any modification to the application source.
Our approach simplifies the task of identifying code regions that can benefit from acceleration and facilitates rapid testing of code modifications on real GPU hardware, thereby making GPUs easily accessible to non-experts.
We evaluated our approach on two proxy applications, four micro benchmarks, and three SPEC OMP 2012 codes with CPU parallelism to demonstrate the simplicity of porting host applications to the GPU.
Our approach closely matched the performance of corresponding manually offloaded kernels, with up to $14.36\times$ speedup on the GPU.
Our evaluation further validates that the \emph{GPU First} methodology can effectively guide porting efforts and identify parallel regions that require reorganization to achieve good scaling behavior on the GPU.
Overall, our proposed approach offers a simple and efficient solution for porting legacy CPU applications to GPUs, enabling non-experts to access GPUs, and facilitating faster development and porting times which will lead to more efficient use of resources.

\section*{Acknowledgement}
This research was supported by the Exascale Computing Project (17-SC-20-SC), a collaborative effort of two U.S. Department of Energy organizations (Office of Science and the National Nuclear Security Administration) responsible for the planning and preparation of a capable exascale ecosystem, including software, applications, hardware, advanced system engineering, and early testbed platforms, in support of the nation's exascale computing imperative.
The views and opinions of the authors do not necessarily reflect those of the U.S. government or Lawrence Livermore National Security, LLC neither of whom nor any of their employees make any endorsements, express or implied warranties or representations or assume any legal liability or responsibility for the accuracy, completeness, or usefulness of the information contained herein.
This work was in parts prepared by Lawrence Livermore National Laboratory under Contract DE-AC52-07NA27344 (LLNL-CONF-827970).
We also gratefully acknowledge the computing resources provided and operated by the Joint Laboratory for System Evaluation at Argonne National Laboratory.

\bibliographystyle{ACM-Reference-Format}
\bibliography{reference}

\end{document}